\documentclass[twocolumn,showpacs,preprintnumbers,amsmath,amssymb]
{revtex4}

\usepackage{dcolumn}
\usepackage{bm}
\usepackage{epsfig}

\def\babar{\mbox{\sl B\hspace{-0.1cm} {\small\sl A}\hspace{-0.1cm} 
\sl B\hspace{-0.1cm} {\small\sl A\hspace{-0.05cm}R}}}

\def\bm{{\it B}}

\begin{document}

\preprint{APS/123-QED}

\title{Report of Snowmass 2001 Working Group E2 :\\
Electron-positron Colliders from the $\phi$ to the $Z$\\}

\author{Zhenguo Zhao}
\affiliation{Beijing Institute of High Energy Physics \\}
\author{Gerald Eigen}
\affiliation{University of Bergen \\}
\author{Gustavo Burdman$^*$}
\affiliation{Boston University \\}
\author{William Marciano}
\affiliation{Brookhaven National Laboratory \\}
\author{David Hitlin}
\affiliation{California Institute of Technology \\}
\author{Mark Mandelkern}
\affiliation{University of California, Irvine \\}
\author{Abi Soffer}
\affiliation{Colorado State University \\}
\author{David Cassel, Lawrence Gibbons}
\affiliation{Cornell University \\}
\author{Klaus Moenig}
\affiliation{DESY, Zeuthen \\}
\author{Joel Butler$^*$, Penelope Kasper, Rob Kutschke, Paul Mackenzie, 
Stephen Pordes, Ron Ray, Tenaji Sen} 
\affiliation{Fermilab \\}
\author{Diego Bettoni, Roberto Calabrese}
\affiliation{University of Ferrara \\}
\author{Caterina Bloise}
\affiliation{Frascati National Laboratory \\}
\author{Daniel Kaplan}
\affiliation{Illinois Institute of Technology \\}
\author{Nobu Katayama, Yasuhiro Okada, Yukiyoshi Ohnishi, Hitoshi Yamamoto$^*$}
\affiliation{KEK\\}
\author{Andrei Gritsan}
\affiliation{Lawrence Berkeley National Laboratory \\}
\author{Steve Dytman}
\affiliation{University of Pittsburgh \\}
\author{Jik Lee, Ian Shipsey$^*$}
\affiliation{Purdue University\\}
\author{Yuri Maravin}
\affiliation{Southern Methodist University \\}
\author{Franz-Joseph Decker, Gudrun Hiller, Peter Kim, David Leith, Sibylle Petrak, 
Steven Robertson, Aaron Roodman, John Seeman}
\affiliation{Stanford Linear Accelerator Center \\}
\author{Marina Artuso, Sheldon Stone}
\affiliation{Syracuse University \\}
\author{Xinchou Lou}
\affiliation{University of Texas, Dallas \\}
\author{Michael Luke}
\affiliation{University of Toronto \\}
\author{Will Johns}
\affiliation{Vanderbilt University \\}
\author{($*$ E2 Working Group Convenor)}
\affiliation{\\}
\date{October 15, 2001 }

\begin{abstract}
We report on the status and plans of experiments now running or proposed
for electron-positron colliders at energies between the $\phi$ and the $Z$.
The $e^{+}e^{-}$ \bm\  and charm factories we considered were PEP-II/\babar,
KEKB/Belle, superKEK, Super\babar, and CESR-c/CLEO-c. We reviewed the programs 
at the $\phi$ factory at
Frascati  and the proposed PEP-N facility at Stanford Linear Accelerator 
Center. We studied the prospects for \bm\  physics with a dedicated linear 
collider $Z$ factory, associated with the TESLA high energy linear collider.
In all cases, we compared the physics reach of these facilities with that
of alternative experiments at hadron colliders or fixed target facilities. 
\end{abstract}

\pacs{Valid PACS appear here}
\maketitle

\section*{Introduction}

In this report we review the status of ongoing and planned electron-positron
collider facilities whose center of mass energies range from the mass of
the $\phi$ meson to that of the $Z$ Boson. In Section 1 and 2, we discuss the 
physics potential of two ``low energy machines'', the $\phi$ factory at 
Frascati and the proposed PEP-N storage ring at Stanford Linear Accelerator. 
Section 3 presents the physics potential of a proposed reorientation of the 
CESR machine and the CLEO detector, known as CLEO-c, which would focus
on topics in charm physics and QCD. In section 4, we discuss the future evolution
of the two asymmetric $e^{+}e^{-}$ \bm-factory facilities, KEKB/Belle and 
PEP-II/\babar\ to superKEK and Super\babar\ and compare their \bm\ 
physics reach to that of existing
and proposed hadron collider experiments. In section 5, we discuss the
potential of a dedicated $Z$ factory associated with a Linear Collider,
in this case TESLA, for \bm\ physics studies and compare its strengths
to those of $e^{+}e^{-}$ and hadron collider experiments. In section 6,
we present our conclusions. This report
is a written version of the E2 Summary Talk 
given at the final plenary session of Snowmass~\cite{shipsey2}.

\section{$\phi$ Factories}

The $\phi$ factory, DA$\phi$NE, at Frascati is a unique facility, in which
electron and positron beams of energy 510 MeV collide~\cite{Bloise}. There are no plans
to build a similar facility elsewhere. While there are several aspects to its
physics program, the E2 working group concentrated on the physics reach of
the KLOE (KLOng Experiment) as compared to planned fixed target Kaon
experiments, which will run at US facilities in the next several years.

\subsection{Status of DA$\phi$NE}

DA$\phi$NE consists of two independent storage rings, one for electrons
of 510 MeV and one for positrons of 510 MeV. The beams intersect at an
angle of 25 milliradians at two locations. The bunch length is 3 cm. The
horizontal bunch size is 2 mm and the vertical size is 0.02 mm. The design
luminosity is  $5\times 10^{32}{\rm cm}^{-2}{\rm s}^{-1}$.

It has been a great challenge to obtain reasonable luminosity. Recently,
a luminosity of $2.5\times 10^{31}{\rm cm}^{-2}{\rm s}^{-1}$ has been achieved.
This is a significant improvement over a year ago and, while still far
below the design, is sufficient to begin to do meaningful physics.
Over the last few months sustained running at  1.3$pb^{-1}$/day has been
achieved. An integrated luminosity of 200$pb^{-1}$ is expected by the
end of calendar 2001.

\subsection{The KLOE Experiment: Description, Goals, and Status}

\begin{figure*}
\begin{center}
\epsfig{figure=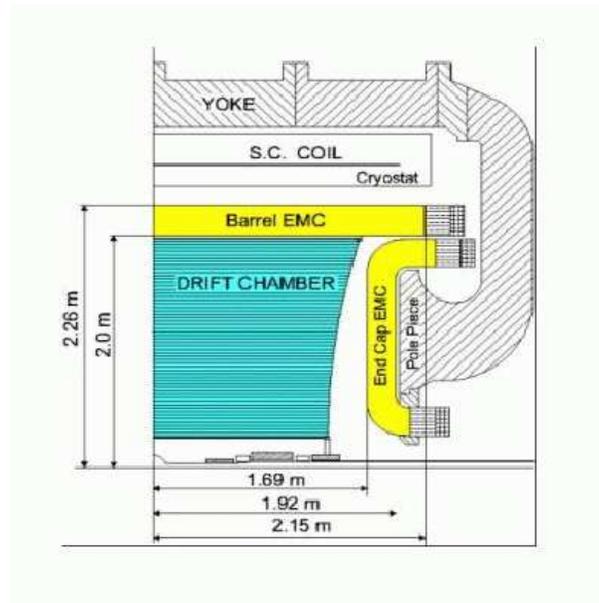,width=8cm,height=8cm}
\caption{A schematic of the KLOE detector}
\label{fig:kloe}
\end{center}
\end{figure*}

\begin{figure*}
\begin{center}
\epsfig{figure=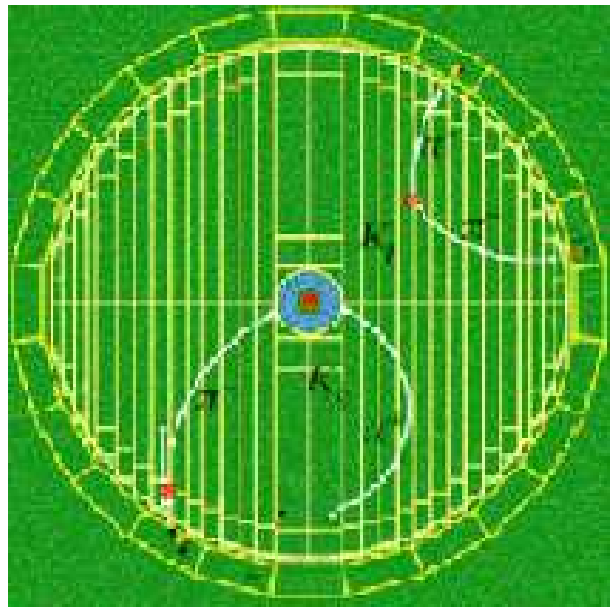,width=8cm, height=8cm}
\caption{A CP violating $K_{L}^{o}$ decay as seen in KLOE}
\label{fig:kloe_event}
\end{center}
\end{figure*}

A main goal of KLOE is to study rare and CP violating decays of the 
$K_{L}^{o}$ mesons which are produced in the decay 
$\phi\rightarrow K_{L}^{o}K_{s}^{o}$.
A schematic of the KLOE detector is given in Fig.~\ref{fig:kloe}.
It has a 5m diameter
superconducting solenoid, which contains a drift chamber and a
lead-scintillator electromagnetic  calorimeter. There is also an endcap 
electromagnetic calorimeter. The drift chamber uses Helium gas to
minimize multiple scattering and $K_{L}^{o}$ regeneration. A CP violating
$K_{L}^{o}$ decay has a very clear signature in the detector, as shown
in Fig.~\ref{fig:kloe_event}.

The physics program of KLOE is quite broad and is described in 
Table~\ref{tab:kloe}.
The table includes physics topics and the approximate luminosity required to 
make meaningful measurments for each topic. It can be seen that some
measurements are already achievable with the current luminosity but the
study of CP violation and rare kaon decays requires significant improvements.

\begin{table}[t]
\begin{center}
\caption[]
{Summary of KLOE Physics Program}
\label{tab:kloe}
\begin{tabular}{|l|c|} \hline
Physics Topic & Integrated Luminosity\\ 
              & ($pb^{-1}$)\\ 
\hline
$\phi$ radiative decays ($f_{o}\gamma, a_{o}\gamma,\eta\gamma,
\eta^{\prime}\gamma) $ & 20-100 \\  
Measurement of $\sigma(\pi\pi)$ (for $g-2$) &  \\ \hline
K semileptonic decays, $Kl4$,               &         \\
$\eta$/$\eta^{\prime}$ mixing, $\ldots$  & 1000 \\ \hline
Tests of CP and CPT violation and  &  \\
measurement of rare K decays & 5000 \\
\hline
\end{tabular}
\end{center}
\end{table}

\subsection{Comparison of Physics Reach of KLOE to Planned Fixed Target
Experiments}

\begin{figure*}
\begin{center}
\epsfig{figure=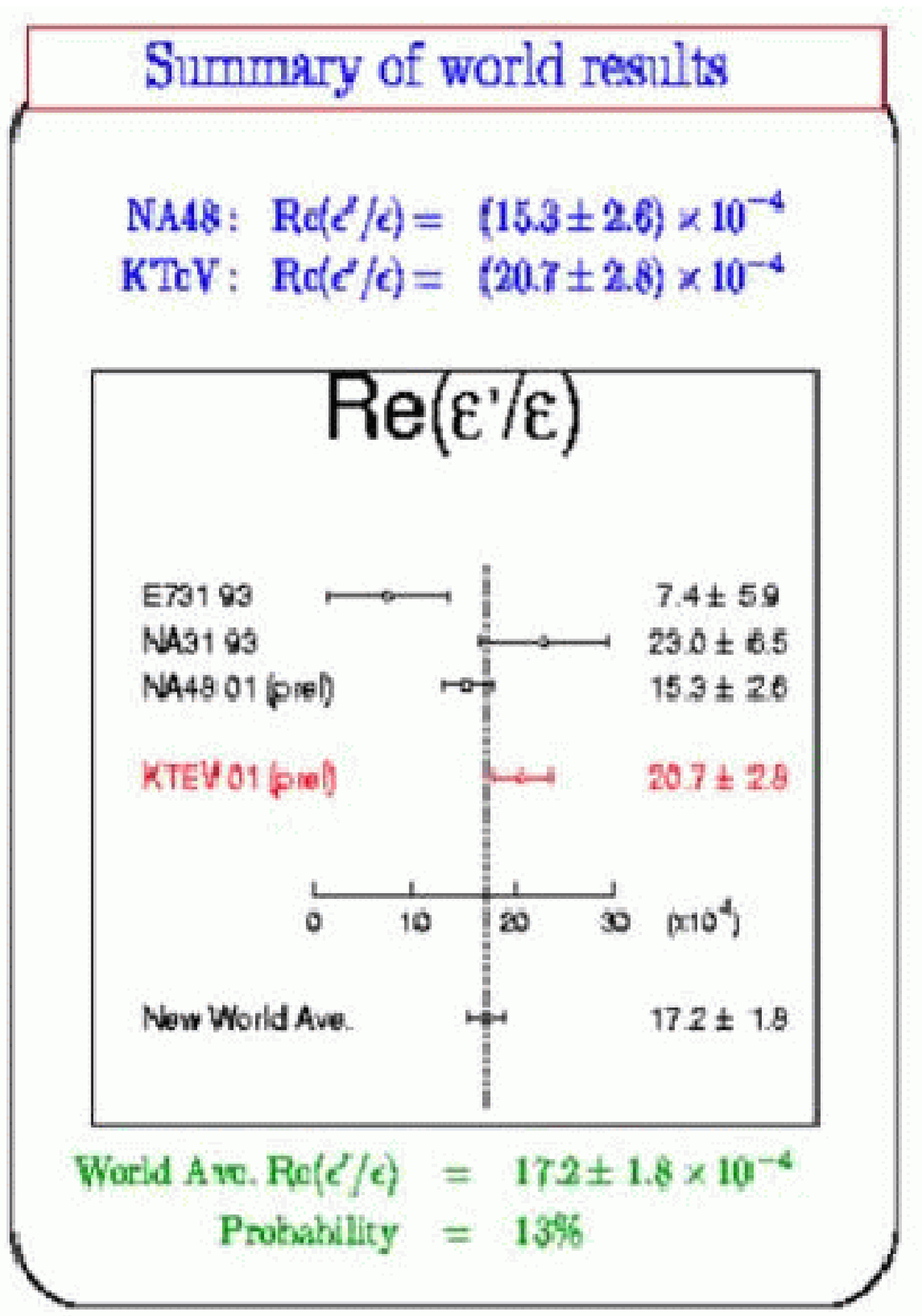,width=7cm,height=8cm}
\caption{World Results on $\frac{\epsilon^{\prime}}{\epsilon}$}
\label{fig:eps}
\end{center}
\end{figure*}

The current status of measurements of ``direct CP violation'' through
the quantity $\epsilon^{\prime}$/$\epsilon$ in Fixed Target Experiments
at CERN(NA48) and Fermilab(KTeV) is shown in Fig.~\ref{fig:eps}. At a 
$\phi$ factory, the double ratio and interferometric methods are complementary
to the Fixed Target experiments. KLOE's goal of 
measuring $\epsilon^{\prime}$/$\epsilon$  to an accuracy of 
$\sim 2\times 10^{-4}$, which requires 5000 $pb^{-1}$, will provide a 
measurement comparable to the other experiments. However, the ability
to extract Standard Model CP parameters from this quantity is, at present,
limited by theoretical uncertainties. 

Another emphasis of future Fixed Target programs in the US is rare kaon
decays, in particular, measurement of the branching fractions  of
\begin{eqnarray}
K^{+}  & \rightarrow & \pi^{+}\nu\bar{\nu}  \\
K_{L}^{o} & \rightarrow & \pi^{o}\nu\bar{\nu}. 
\end{eqnarray}
The first of these provides a measurement of $V_{td}$ and the second
is a direct indicator of the CKM parameter $\eta$. The branching fractions
are very small, of order a few$\times 10^{-11}$. Very high kaon fluxes
are needed and Fixed Target experiments that want to detect them must 
withstand formidable backgrounds and run at very high rates.

The $\phi$ factory has very desirable features for doing these measurements
which avoid many of the problems of the Fixed Target experiments. However,
even with 5000 $pb^{-1}$, only about $10^{10}$ $K_{L}K_{s}$ pairs will
be produced so the Standard Model expectations cannot quite be reached.
The branching fraction for the now observed decay 
$K^{+}  \rightarrow \pi^{+}\nu\bar{\nu}$ is already too low for KLOE to reach.
However, if there is new physics, outside the Standard Model, in the
decay $K_{L}^{o}\rightarrow \pi^{o}\nu\bar{\nu}$, which currently has a 
limit only of order $10^{-6}$,
this process could be within the range of the KLOE experiment.
Thus, KLOE has a few year window to push the sensitivity of 
$K_{L}^{o}\rightarrow \pi^{o}\nu\bar{\nu}$ in the hope that new physics
might be present there. If the Standard Model processes are the dominant
ones, then ultimately this decay will have to be observed in Fixed Target
kaon experiments. See ~\cite{Ray} for further details.

\begin{figure*}
\begin{center}
\epsfig{figure=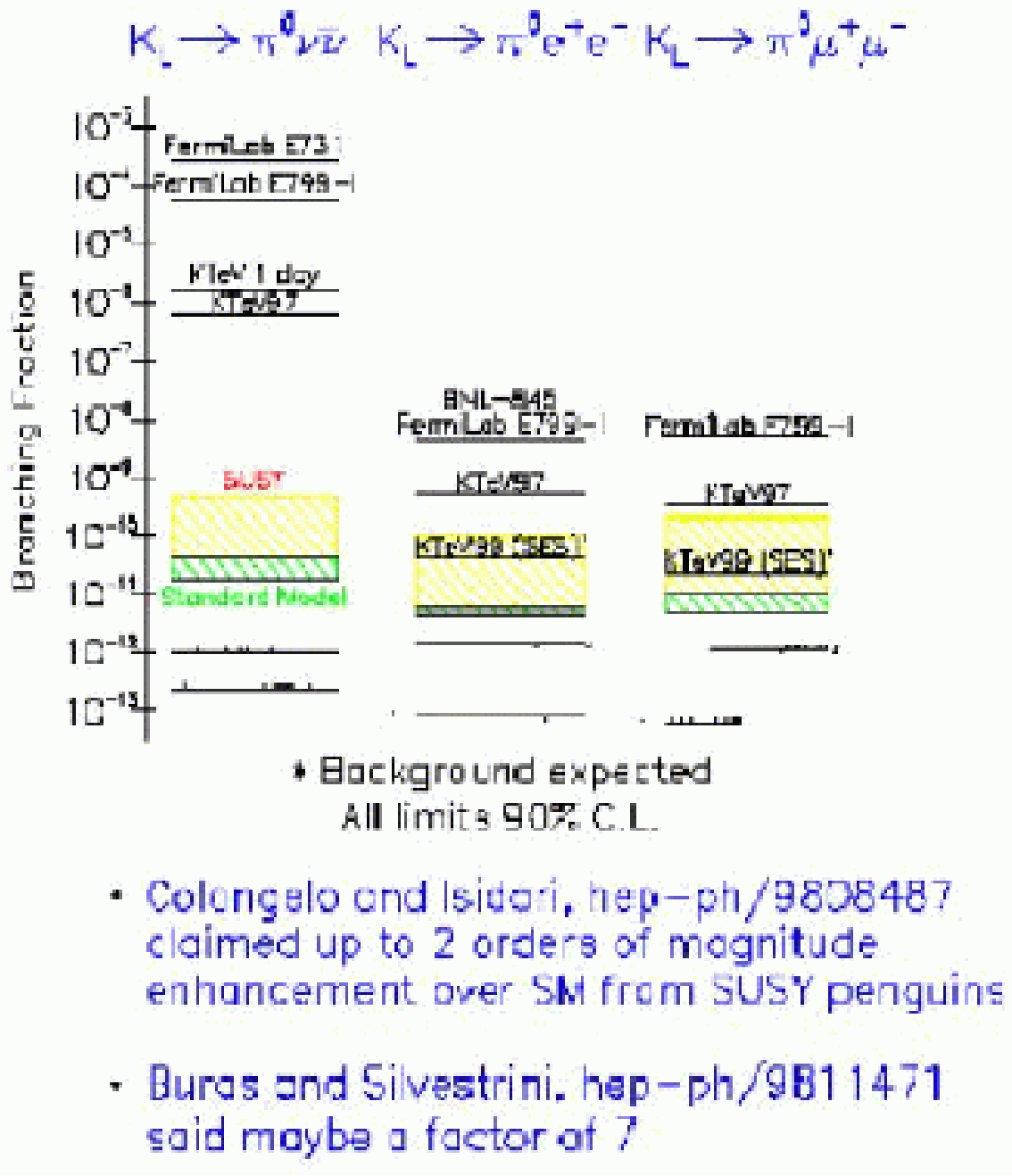,width=8cm,height=8cm}
\caption{Current and expected results on rare K decays. For each mode, the 
two lines corresponding to the greatest sensitivity are for 
the Kopio experiment ($K_{L}^{o}\rightarrow \pi^{o}\nu\bar{\nu}$) and 
the KAMI proposal (all three modes). Note KAMI is not approved. }
\label{fig:rare_k_decay}
\end{center}
\end{figure*}

\section{PEP-N}

\begin{figure}
\begin{center}
\epsfig{figure=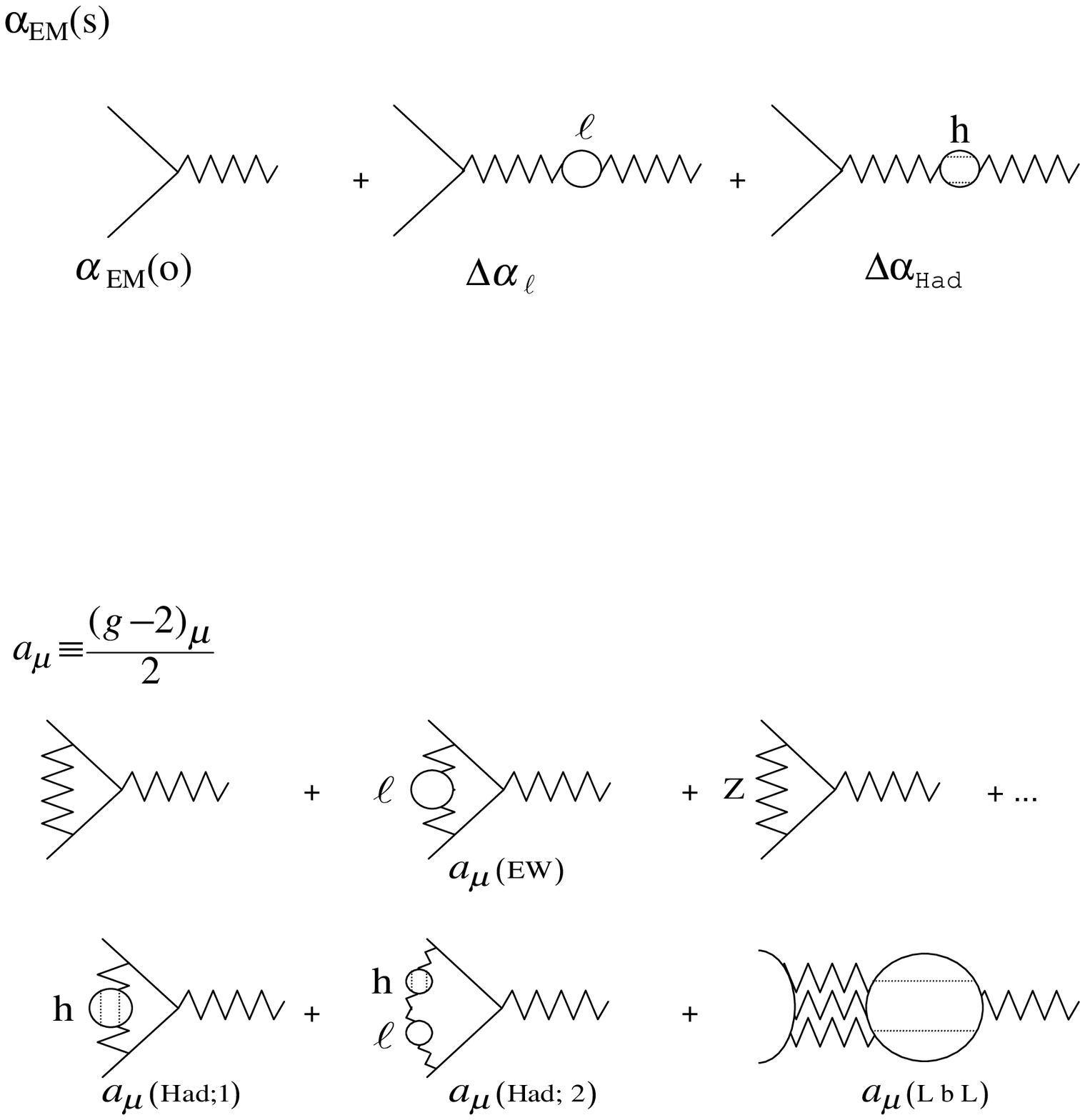,width=6cm,height=9cm}
\caption{Feynman diagrams for radiative corrections to $\alpha_{em}$ and
$(g-2)_\mu$}
\label{fig:diagrams}
\end{center}
\end{figure}

PEP-N is a proposed novel extension of PEP-II. The machine  
is an asymmetric collider
consisting of the PEP-II Low Energy Ring (LER) (3.1 GeV) 
and a new electron storage
ring (Very Low Energy Ring, VLER) of energy $100\ {\rm MeV} < E_e < 800\ {\rm MeV}$.
The accessible center of mass (CM) energy is 
$1.2\ {\rm GeV} < \sqrt{s} < 3.15\ {\rm GeV}$. This machine would 
run simultaneously with PEP-II operation at the $\Upsilon(4S)$.

There is a rich variety of important physics measurements that are accessible
at this collider. The most prominent are the high-precision measurement
of the ratio, R \cite{mark}\cite{marciano}, of the hadron total cross
section to the muon pair cross section and the determination of nucleon form
factors \cite{roberto}. 
Other physics topics which can be studied at PEP-N include meson form
factors, vector meson spectroscopy, the search for non $q\overline{q}$
states and $\gamma\gamma^*$ interactions. 

In our view the most important single measurement that PEP-N could 
contribute is the determination of R with greatly improved precision.
In this report we will focus solely on the physics motivation
and challenges of measuring R.

\subsection{The Measurement of R}

\begin{figure}[t]
\begin{center}
\epsfig{figure=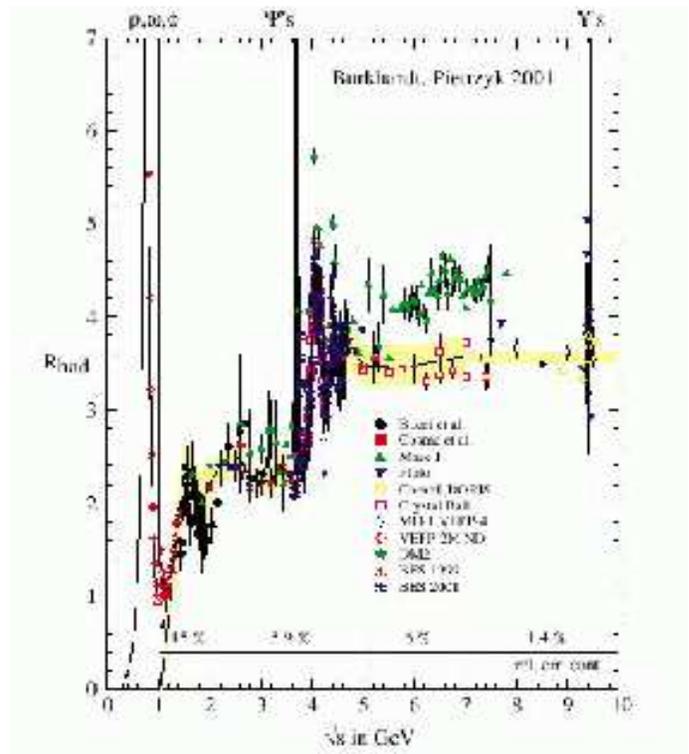,width=9cm,height=10cm}
\caption{R$_{had}$ including resonances with the parameterization of
Burkhardt and Pietrzyk.}
\label{fig:R}
\end{center}
\end{figure}

Testing the consistency of the Standard Model requires a variety of 
measurements for which radiative corrections play a crucial role. Two of the most
important examples are (a) Higgs mass bounds from precision measurements at LEP
and electroweak natural
relations (i.e. the evolution of $\alpha$ to the $Z$ pole), and 
(b) Interpretation of the BNL $g_\mu-2$ experiment \cite{E821}. In addition, 
future higher precision
experiments, such as Giga-$Z$, will depend on radiative corrections 
being precisely known.

The parameters of the electroweak model can be taken as $G_F$,
$\alpha_{em}(0)$, $M_Z$, $m_H$ and the fermion masses and mixings. In order to compute
physical quantities we must include radiative corrections which renormalize
charges, masses and magnetic moments as shown in Fig. \ref{fig:diagrams}.
Although the electroweak radiative corrections are calculable, the hadronic
radiative corrections are not. However the lowest-order hadronic radiative
corrections can be obtained from $e^+ e^-\rightarrow hadrons$
using dispersion relations and unitarity. The forward scattering amplitude
for virtual photons interacting with the vacuum is related to the total
cross section for that process by the Optical Theorem.

\subsubsection{ The evolution of $\alpha$ to $M_Z$}

\begin{figure}[t]
\centering
\epsfig{figure=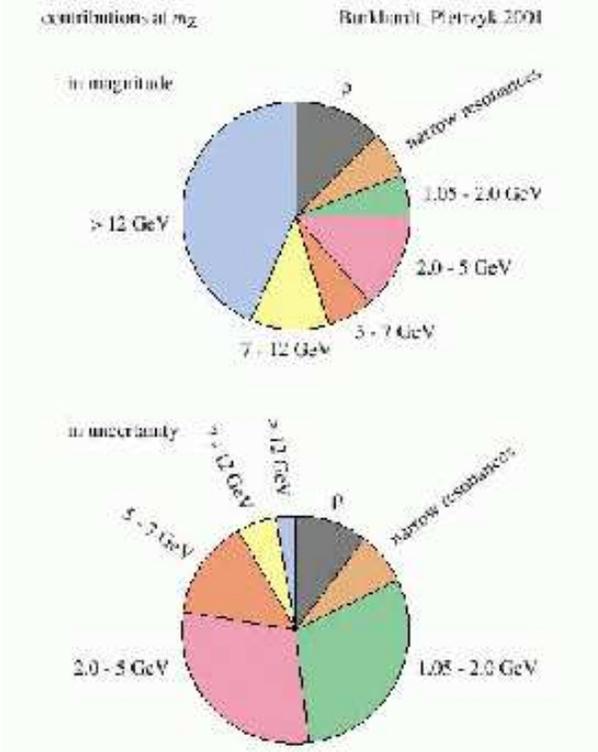,width=8cm,height=10cm}
\caption{Relative contributions to $\Delta\alpha^{(5)}_{had}(M_Z^2)$ in
magnitude and uncertainty from Burkhardt and Pietrzyk.}
\label{fig:pie}
\end{figure}

In leading order perturbation theory:

\begin{eqnarray}
\Delta \alpha(s)&=&\frac{\alpha}{3\pi}
\sum_{m_f^2<<s}Q_f^2N_{cf} (ln\frac{s}{m_f^2}-\frac{5}{3}) \nonumber \\
&=&\Delta\alpha_{leptons}(s)+ \Delta\alpha_{hadrons}(s)
\end{eqnarray}
This expression is inadequate for the hadronic contribution, which can be 
obtained from the measurement of R. For $(2m_t)^2>>s>>(2m_b)^2$ we have:

\begin{equation}
\Delta \alpha(s)= 
\Delta\alpha_{leptons}(s)+\Delta\alpha_{hadrons}^{(5)}(s)
\end{equation}

\begin{equation}
\Delta\alpha_{hadrons}^{(5)}(s)=-\frac{\alpha 
s}{3\pi}\int_{4m_\pi^2}^{\infty}
\frac{R(s')}{s'(s'-s)}ds'
\end{equation}

\noindent Our current knowledge of R below 10 GeV is shown in 
Fig. \ref{fig:R}. $\Delta\alpha(M_Z^2)$ is
of particular importance for predicting the $W$ mass and $Z$-pole
asymmetries and has been calculated by many authors including 
Burkhardt and Pietrzyk (BP) \cite{BP}. BP find
$\Delta\alpha_{hadrons}^{(5)}(M_Z^2)=0.02761\pm 0.00036$ (1.3\%)  
corresponding to $1/\alpha(M_Z^2)= 128.936\pm 0.046$ (0.037\%). 
The largest contributions to the uncertainty in
$\Delta\alpha_{hadrons}^{(5)}(s)$ are from the measured values of R in
the regions 1.05$<\sqrt{s}<$2.0 GeV and 
2.0$<\sqrt{s}<$5.0 GeV, each contributing about
0.8\% as shown in Fig. \ref{fig:pie} from Ref.
\cite{BP}.  The latter uncertainty decreased significantly
after inclusion of the recent BES (inclusive) data \cite{BES}, even though
the measurements between 2 and 3 GeV have large errors and potentially
significant systematic uncertainties. 
The uncertainties in the contributions from different intervals are systematics
dominated. However BP combines the errors in
quadrature. If one were to sum the systematic errors, the uncertainty would
be 3\%.

As noted in ~\cite{mark}, the consistency of R measurements between 3 and 4 GeV and between 5 
and 8 GeV is poor. Absolute cross 
sections are difficult to measure and there may be significant 
systematic errors in the measurements beyond those estimated by the 
experiments.   

$\Delta \alpha (M_Z^2)$ enters in electroweak physics via
\begin{equation} 
\sin^2\Theta\cos^2\Theta=\frac{\pi\alpha}{\sqrt 2 
G_F M_Z^2}\frac{1}{1-\Delta r}
\end{equation}
where
\begin{equation}
\Delta r=\Delta\alpha(M_Z^2)-f(\sin^2\Theta)\delta \rho +\Delta r_{Higgs} 
+ \Delta 
r_{other}
\end{equation}
and
\begin{equation}
\delta \rho \simeq \frac{\sqrt 2 G_F}{16 \pi^2}3 m_t^2
\end{equation}
\begin{eqnarray}
\Delta r_{Higgs} &\simeq& \frac{\sqrt 2 G_F M_W^2}{16 
\pi^2}\{c^H(\sin^2\Theta)(ln\frac{m^2_H}{M^2_W}-\frac{5}{6})\} ; \nonumber \\
&&m_H>>M_W
\end{eqnarray}
$c^H(\sin^2\Theta)$ and $f(\sin^2\Theta)$ are dependent on the definition
of $\sin^2\Theta$, i.e. the renormalization method. In the on-shell
scheme, for example, $C_W^H=11/3$ and $f_W(\sin^2\Theta)=\cot^2\Theta_W
\simeq 3.35$.

\begin{figure}[t]
\centering
\epsfig{figure=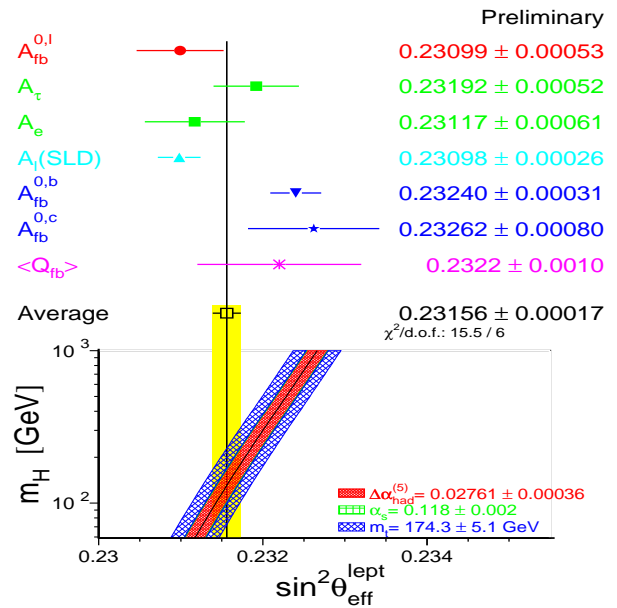,width=8cm,height=8cm}
\caption{Measurements of the effective leptonic $\sin^2\theta_W$ and the
predictions of the Standard Model with uncertainties due to 
$\Delta\alpha^{(5)}_{had}(M_Z^2)$ and $m_t$.}
\label{fig:lepewwg}
\end{figure}

The resulting fractional theoretical uncertainty in $M_W$ is $\sim
0.23 \delta \Delta \alpha$. The contribution from the 0.0004 uncertainty
in $\Delta\alpha_{hadrons}^{(5)}(s)$ is about 75 MeV, compared to the
experimental uncertainty of 56 MeV. 
Measurements of the effective leptonic $\sin^2\theta_W$ and the
predictions of the Standard Model with uncertainties due to 
$\Delta\alpha^{(5)}_{had}(M_Z^2)$ and $m_t$ from the LEPWG \cite{LEPEWWG} 
are shown in Fig. \ref{fig:lepewwg}. 

\begin{figure}[t]
\centering
\epsfig{figure=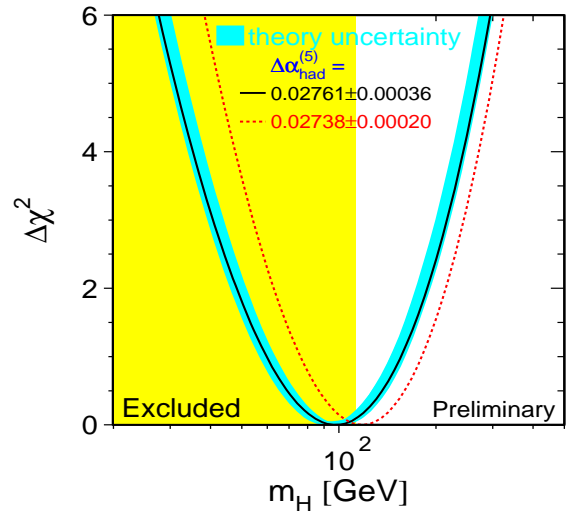,width=8cm,height=8cm}
\caption{Light Higgs mass prediction of precision electroweak data, with
uncertainty due to hadronic corrections.}
\label{fig:blueband}
\end{figure}

The effective weak mixing angle, can be determined from
$Z$-pole asymmetry data, etc. without knowledge of the top and Higgs masses.
The Standard Model prediction is given as a function of $m_H$ with
uncertainties due to $\Delta\alpha_{hadrons}^{(5)}$, $m_t$, and $m_Z$. The
uncertainty in $sin^2\Theta^l_{eff}$ due to $\Delta\alpha_{hadrons}^{(5)}$
is $\sim sin^2\Theta^l_{eff}\Delta\alpha_{hadrons}^{(5)}\sim\pm 0.0001$,
that due to $m_t$ is also about 0.0001, and that due to $M_Z$ $<<0.0001$,
compared to the experimental error of $0.00017$. The overall fit to $m_H$ from
all electroweak data, shown in Fig.\ref{fig:blueband}, yields an estimate
of $\sim 100^{+57}_{-38}$GeV where the dominant contribution to the
uncertainty, $\sim 20$ GeV, is from $\Delta\alpha_{had}^{(5)}$.

\subsubsection{$(g-2)_\mu$}

We now consider hadronic corrections to the muon magnetic moment.
The Standard Model prediction for $a_\mu \equiv(g-2)_\mu/2$ is:

\begin{equation}
 a_\mu(theory)=a_\mu(EW)+a_\mu(Had).
\end{equation}
$a_\mu(EW) \equiv a_\mu(QED)+a_\mu(Weak)$ is calculable to a few
parts in $10^{11}$. The uncertainty in $a_\mu$ is dominated by that in
$a_\mu(Had)$ which is usually broken up into the leading vacuum
polarization contribution $a_\mu(Had;1)$ of order
$(\frac{\alpha}{\pi})^2$, the higher order vacuum polarization
contribution $a_\mu(Had;2)$ of order $(\frac{\alpha}{\pi})^3$, and the
hadronic light-by-light contribution $a_\mu(LbL)$, also of order
$(\frac{\alpha}{\pi})^3$. The first of these is related to R by a
dispersion relation, and the second and third must be estimated.

\begin{equation}
a_\mu(Had;1)=(\frac{\alpha_{em}
m_\mu}{3\pi})^2\int^\infty_{4m_\pi^2}\frac{ds}{s^2}K(s)R(s)
\end{equation}
where

\begin{eqnarray}
K(s) &=& \frac{3s}{m_\mu^2}\{x^2(1-\frac{x^2}{2}) \nonumber \\
     &+&(1+x)^2(1+\frac{1}{x^2})\{ln(1+x)-x+\frac{x^2}{2}\} \nonumber \\
     &+& \frac{1+x}{1-x}x^2ln x\} 
\end{eqnarray}
with

\begin{equation}
x=\frac{1-\beta}{1+\beta}, \beta=\sqrt{1-\frac{4m_\mu^2}{s}}.
\end{equation}
Note the weighting of $R(s)$ is $1/s^2$, making the low energy regime more
important than for $\alpha(s)$. Some recent analyses have used $\tau$ decay
data to supplement $e^+ e^-$ data.  Here CVC is used to relate processes
through the vector charged weak current to comparable processes through the
isovector E.M. current assuming no second class weak currents, which implies
that the contribution of the axial vector current to G+ decays is zero. Thus
annihilation cross sections with $G=C(-1)^I=+1$ (G+, i.e. $n_\pi$ even) are
obtained from the rates of corresponding $\tau$ decays. While $\tau$ decay
data is useful at the current level of accuracy, I-spin violation and
effects such as initial and final state radiation must be understood if we
are to rely on it at smaller experimental errors, as emphasized by Eidelman
and Jegerlehner \cite{EJ,Jeger1}and by Melnikov \cite{Melni}.
PQCD is used at energies$>12$ GeV by all authors because of the lack of
data. The result of Davier and Hocker (DH) \cite{DH}, who use QCD sum rule
constraints at low energy as well as $\tau$ data, is $a_\mu(Had;1)=6924(62)
\times 10^{-11}$, giving the dominant uncertainty in $a_\mu$. The more
conservative result of Jegerlehner is 6987(111).

The higher order hadronic vacuum polarization and hadronic
light-by-light contribution to $a_\mu$ are comparable. However while the
uncertainty in the former is several parts in $10^{11}$, the uncertainty in
the latter is much larger. The detailed calculations done by Hayakawa and
Kinoshita \cite{HK} and by Bijkens, Pallante and Prades \cite{BPP}
give a negative $a_\mu^{LbL}$~\cite{Knecht}.
 Marciano and Roberts in their recent review
\cite{MR} combine in quadrature the DH result for
$a_\mu(Had;1)= 6924(62) \times 10^{-11}$ and $a_\mu(LbL)=-85(25)\times
10^{-11}$(the average of HK and BPP taking the average of the quoted
uncertainties) for an overall result of $a_\mu^{SM}=116 591 597(67) \times
10^{-11}$. This is  to be compared with the BNL E821 \cite{E821} result of $116
592 020(160)  \times 10^{-11}$. The discrepancy is $423(173)\times
10^{-11}$~\cite{Knecht}. Other authors regard the light-by-light calculation as
model-dependent and less reliable~\cite{mark}.  BNL E821 ultimately 
anticipates an uncertainty of $40 \times 10^{-11}$. Clearly improved 
knowledge of $a_\mu(Had;1)$ and $a_\mu(LbL)$ are required to exploit 
high-precision measurements of $(g-2)_\mu$. The former will greatly 
benefit from better $e^+ e^-$ data below 3 GeV.

\subsection{Experimental Requirements}

\begin{figure*}[t]
\centering
\includegraphics*[width=.45\textwidth]{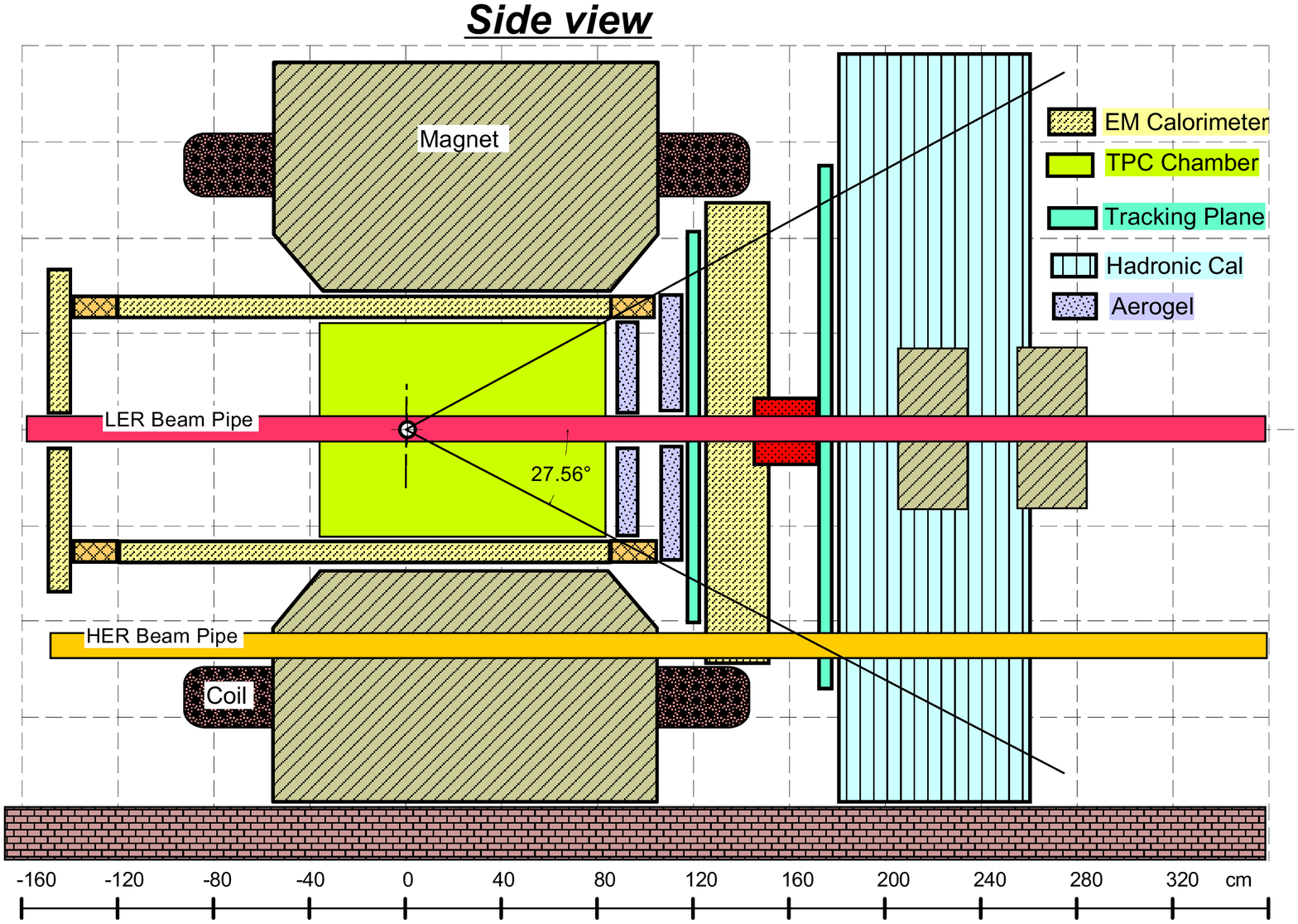}  
\includegraphics*[width=.45\textwidth]{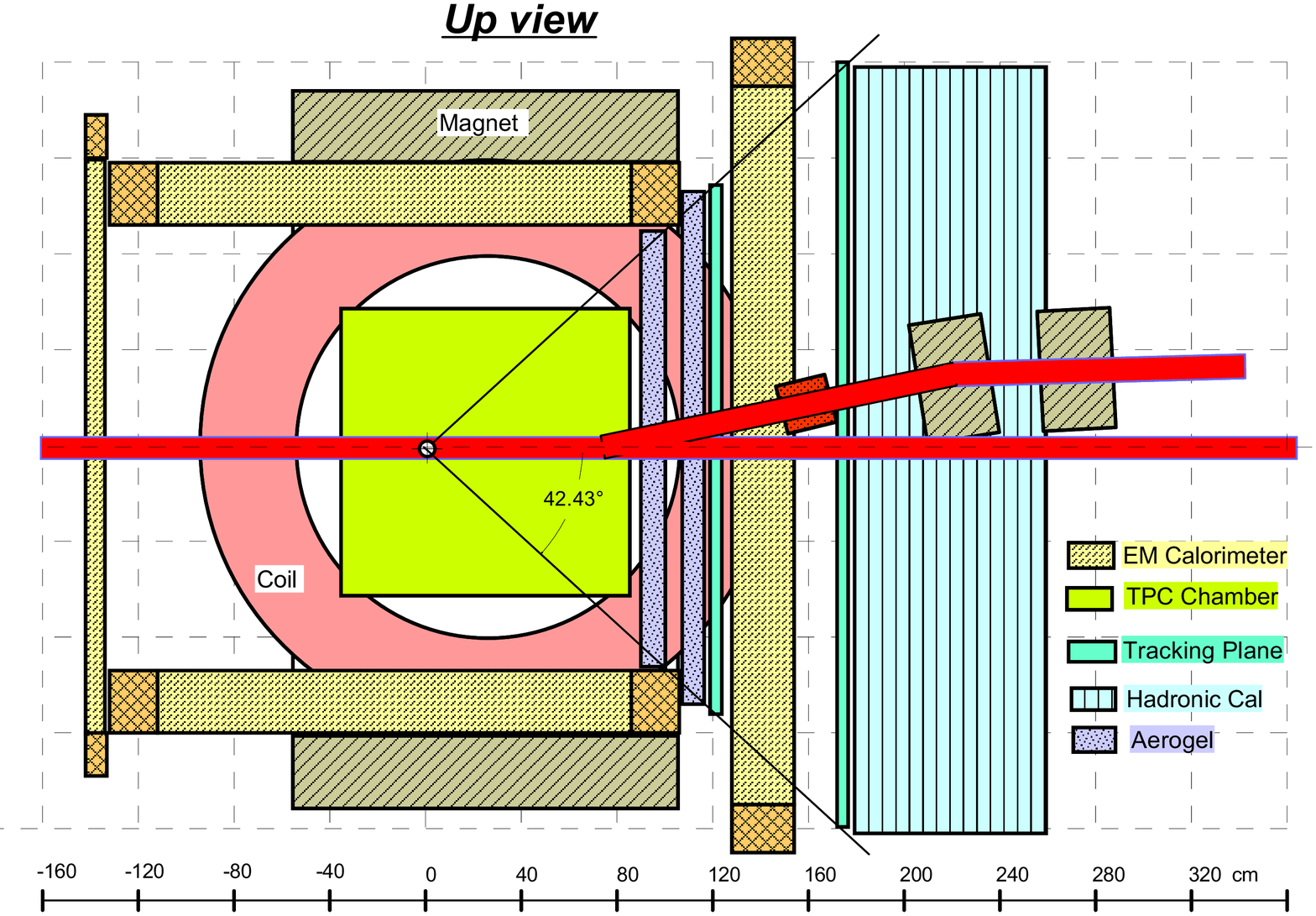}
\caption{PEP-N detector layout: side view (left) and top view (right).} 
\label{fig:layout}
\end{figure*}

Two methods can be used to measure R:
\begin{itemize}
\item {\bf Inclusive approach:} hadronic events are
defined inclusively by requiring a minimum number of particles in the 
detector. 
In order to measure the cross section $\sigma (e^+e^- \rightarrow hadrons)$ 
the acceptance is required. Due to the large number 
of contributing channels, a Monte Carlo simulation is used, 
leading to potentially large systematic errors and rendering this method 
unsuitable for a high-precision (1-2 \%) measurement of R. 
\item {\bf Exclusive approach}: the cross section of each individual channel 
contributing to R is measured. Events must be completely reconstructed
with high efficiency, and acceptances for each channel must be well known. 
With this method an accuracy of 1-2 \% in R can be
reached, as shown by the recent VEPP-2M measurements. 
\end{itemize}

To measure R with a precision of the
order of 2 \% (or better), the PEP-N experiment is designed to
use the exclusive method. The detector has a large acceptance
and is able to measure the absolute position of charged and neutral particles.
In addition, since  
$\sigma(e^+e^- \rightarrow n\overline{n})$ is a sizeable fraction
of the total hadronic cross section (e.g. 2.5 \% at $\sqrt{s} = 2\ {\rm GeV}$),
$n\overline{n}$ detection capability is needed. 

The proposed PEP-N detector must satisfy the following requirements:

\begin{itemize}
\item {\bf Low mass tracking.} In the energy range of 
PEP-N multiple scattering
contributes significantly to the momentum resolution ($\approx 2 \%$);
\item {\bf Momentum measurement with good accuracy.} A high-precision 
measurement of R requires the ability to reconstruct efficiently every 
individual final state. This can be 
done by means of topological selections
and kinematic fitting. The ability
to identify each channel contributing
to R depends crucially 
on a high-precision measurement of the
momenta.
\item {\bf Electromagnetic (EM) calorimetry.} 
The EM calorimeter provides 
the direction and energy of photons with high
precision and accuracy down to 100 MeV or below, and
identifies Bhabhas used for the luminosity
measurement. 
\item{\bf Particle ID} is necessary for $\pi/K$ separation; 
this feature
is crucial to distinguish between and reconstruct efficiently 
final states containing pions and kaons.
\item{\bf Luminosity measurement} with an accuracy of the 
order of 1 \% or better.
\item{\bf $n\overline{n}$ capability}
\end{itemize}

As PEP-N is an asymmetric machine, the CM is travelling at 
$0.6 < \beta_{CM} < 0.94$. 
In consequence, slow 
particles in the CM frame are boosted 
to momenta ranging from a few hundred 
MeV to 1-2 GeV, simplifying detection and reducing the angular coverage
needed to obtain full acceptance.  
The asymmetric operation has the additional advantage  
of simplifying beam separation.

Another important feature of the PEP-N design 
is the magnet. The magnetic
field required to perform beam separation with minimal interference
with PEP-II operation is a 
weak dipole field ($B \approx 0.3\ T$). This
field is also used by the experiment for the measurement of
charged particle momenta. Therefore, the tracking system is housed 
inside the magnet gap which, as a consequence, has to be made big enough
to give a suitable acceptance. Considerable effort has been expended 
to design a magnet with a sufficiently uniform field.

Assuming an average instantaneous luminosity 
of $5 \times 10^{30} {\rm cm}^{-2}{\rm s}^{-1}$
and a detection efficiency of 50 \%, the expected 
hadronic event rate for the
measurement of R is 10,000 events per day.
A 1-2 day data taking period at each CM energy 
provides statistical accuracies better than 1 \%.
PEP-N plans to take data at intervals of 10 MeV.
Several hundred days
of data taking are required 
to cover the energy region between 1.2 GeV and 3.15 GeV.

Taking a maximum total cross section of 100 nb 
and maximum instantaneous
luminosities of $10^{31} {\rm cm}^{-2}{\rm s}^{-1}$, the 
event rate (excluding 
backgrounds) is 1 Hz. Backgrounds will increase this rate
but should present no problem for the detector.

The proposed PEP-N detector layout is shown in fig. \ref{fig:layout}. 
The central detector is housed inside the magnet gap. It
consists of a time projection chamber (TPC) using a slow He based gas 
providing $\sigma = 200-300 \mu{\rm m}$ and dE/dx capability. It is 
proposed to use GEMs for 
the readout to eliminate the $E \times B$ term. The EM 
calorimeter modules are 
located outside the magnet. Energy resolution of a few percent 
down to 100 MeV and good time resolution can be achieved with a lead and
scintillating fiber technology based on the KLOE design.
Particle ID is achieved with two 10 cm thick KEDR style
aerogel counters, which achieve $4 \sigma ~\pi/K$
separation between 600 MeV/c and 1.5 GeV/c.
The hadron calorimeter design was not chosen at the time of 
writing this report. A scintillator based calorimeter
or an extension in depth of the EM 
calorimeter were under investigation.
The dipole magnet and 
the central detector are not centered on the interaction
point. They are shifted 25 cm in the forward direction
to increase the
path inside the magnetic field for particles produced
in the forward direction.

The forward detector consists of two 
silicon aerogel counters for particle ID,
additional tracking planes (drift chambers) 
as well as EM and hadronic 
calorimeter modules. 
Also shown in fig. \ref{fig:layout} are the HER (High Energy Ring), LER
and VLER beam pipes.

The proposed  schedule for PEP-N is as follows. A proposal review is planned 
for summer of 2001. If approval is granted, then
in 2003 the injector gun, linac, and transport lines would be installed. Also modifications to the
PEP-II LER and HER would be made. The first injector beam test would be in October 2003. In summer 2004, the VLER ring,
detector magnet, and detector would be installed. In October 2004, 
first VLER injected beam tests are foreseen.
In January 2005, first collisions would occur.

\subsection{Summary}

The determination of R in this energy range is of particular importance and is timely.
The statistical error achievable is negligible. However, there was
no clear demonstration that the required systematic error of about 2\% 
(dominated by knowledge of the acceptance) is achievable. Studies stimulated 
by the E2 group are ongoing to address this concern.
In one approach, a CLEO-c $10^9$ {\it J}/$\Psi$ run would 
yield precision  {\it J}/$\Psi$ absolute
branching ratios, which could be used by PEP-N in a calibration run at the 
{\it J}/$\Psi$ for a precision determination 
of the acceptance.  The PEP-N detector design appears to be sound.
There is no new technology except for the GEM readout of the TPC.  
We conclude that the physics program of PEP-N is well defined,
important and unique and the required number of events can be
obtained  in five years. However, control
of systematic errors needs to be carefully evaluated before proceeding.

\section{Charm Physics with CLEO-$\lowercase{\rm c}$}

For many years, the CLEO experiment at the Cornell Electron Storage
Ring, CESR, operating on the $\Upsilon$(4S) resonance, has provided much of the
world's information about the $B_{d}$ and $B_{u}$ mesons. 
At the same time, CLEO, using the copious continuum pair production at the
$\Upsilon$(4S) resonance
has been a leader in the study of charm and $\tau$ physics. Now that the
asymmetric \bm-factories have achieved high luminosity, CLEO is
uniquely positioned  to advance
the knowledge of heavy flavor physics by carrying out several measurements
near charm threshold, at center of mass energies in the 3.5-5.0 GeV region.
These measurements address crucial topics which benefit from the high 
luminosity and experimental constraints which exist near threshold but
have not been carried out at existing charm factories because the luminosity 
has been too low, or have been carried out previously with meager statistics.
They include:  
\begin{enumerate}
\item Charm decay constants $f_D, f_{D_s}$; 
\item Charm absolute branching fractions;  
\item Semileptonic decay form factors; 
\item Direct determination of $V_{cd}$ \& $V_{cs}$;  
\item QCD studies including: \\
Charmonium and bottomonium spectroscopy; \\ 
Glueball and exotic searches; \\ 
Measurement of R between 3 and 5 GeV, via scans; and \\ 
Measurement of R between 1 and 3 GeV, via ISR (Initial State Radiation).  
\item Search for new physics via charm mixing, {\it CP} violation
and rare decays; and 
\item $\tau$ decay physics.
\end{enumerate}

The CLEO detector can carry out this program with only minimal modifications.
The CLEO-c project is described at length in \cite{cleo-c}.
It was also described in several talks at this workshop: 
\cite{artuso1} - \cite{maravin}. Theoretical issues in charm physics
were covered in talks \cite{Mackenzie} - \cite{Brodsky2}.
A very modest upgrade to the storage ring, described elsewhere
in these proceedings, is required to achieve the required luminosity.  
Below, we summarize the advantages of
running at charm threshold, the minor modifications required to optimize
the detector, examples of key analyses, a description of the proposed run
plan, and a summary of the physics impact of the program.

\subsection {Advantages of running at charm threshold}

\begin{figure}[t]
\begin{center}
\epsfig{figure=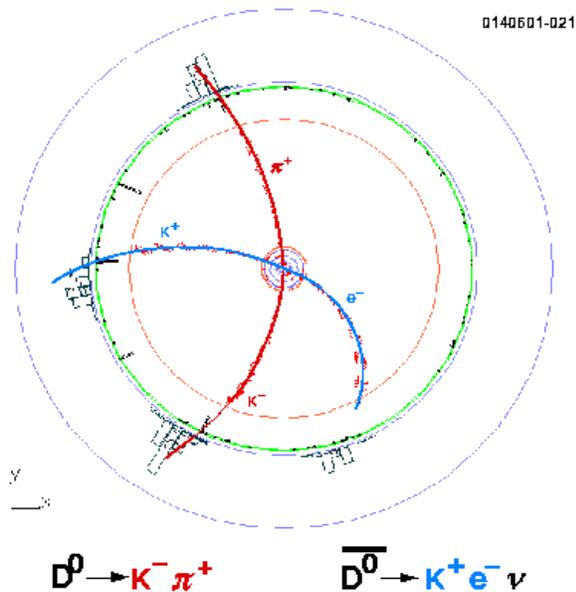,width=8cm,height=8cm}
\caption {A doubly tagged  event at the  $\psi(3770)$}
\label{fig:cleoc_event}
\end{center}
\end{figure}

The \bm-factories, running on the $\Upsilon$(4S) 
will have produced 500 million 
charm pairs from the underlying continuum by 2005. However, there are 
significant advantages of running at 
charm threshold:
\begin{enumerate}  
\item Charm events produced at threshold are extremely clean; 
\item Double tag events, which are key to making absolute branching fraction 
measurements, are pristine;
\item Signal/Background is optimum at threshold;
\item Neutrino reconstruction is clean; and
\item Quantum coherence aids $D$ mixing and {\it CP} violation studies 
\end{enumerate}
These advantages are dramatically illustrated in Figure~\ref{fig:cleoc_event}, 
which shows a picture of a simulated and fully reconstructed 
$\psi(3770)\to D\bar{D}$ event.

\begin{figure}[t]
\begin{center}
\epsfig{figure=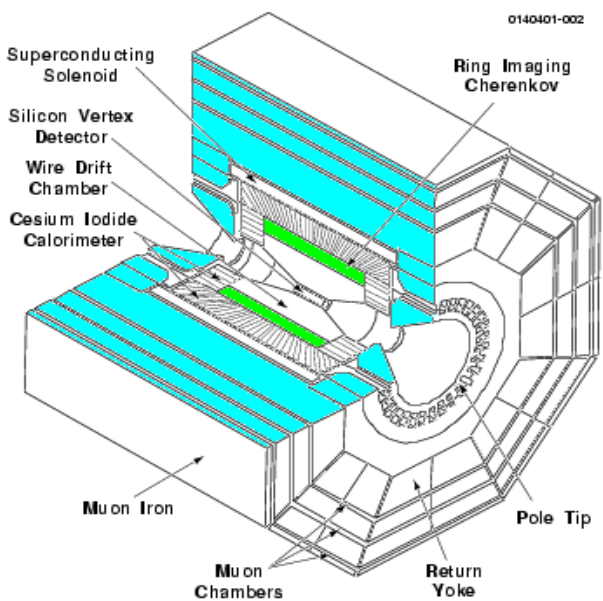,width=8cm,height=8cm}
\caption {The CLEO III detector}
\label{fig:cleo3_det}
\end{center}
\end{figure}

\begin{figure*}
\begin{center}
\epsfig{figure=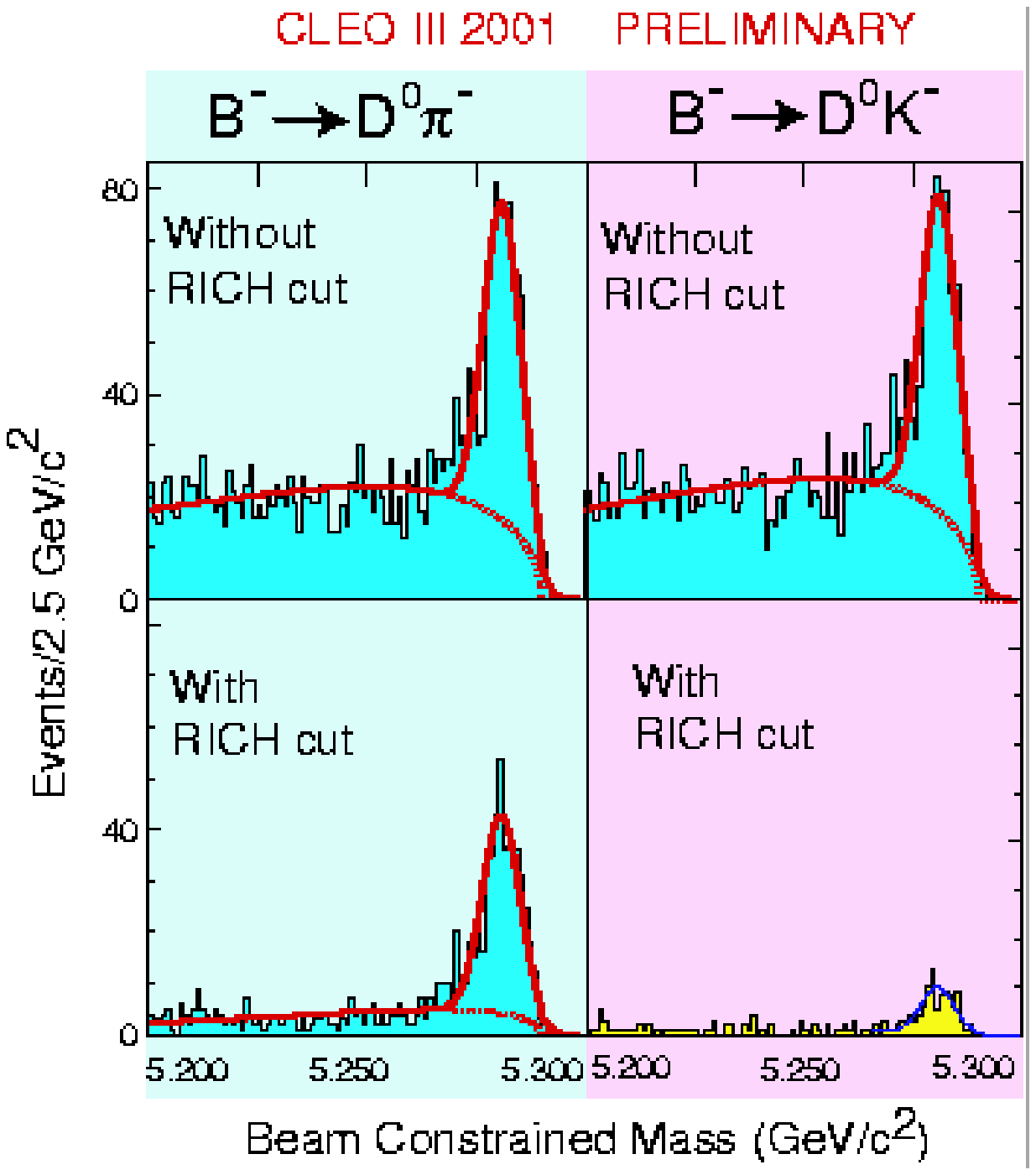,width=8cm,height=8cm}
\epsfig{figure=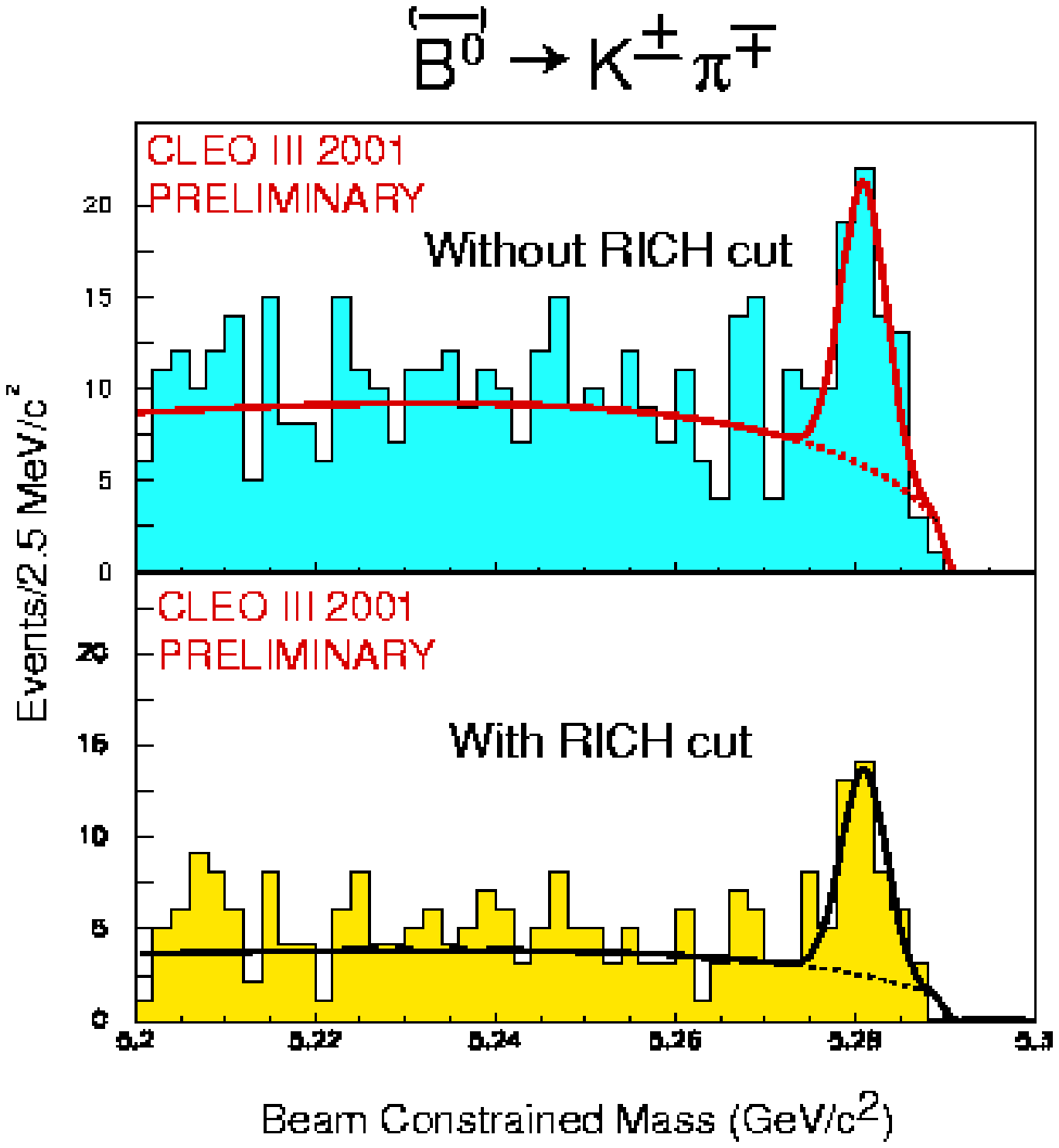,width=8cm,height=8cm}
\caption 
{(Left) Beam constrained mass for the Cabibbo allowed decay 
$B\to D\pi$  and the 
Cabibbo suppressed decay $B\to DK$ with and without RICH information. 
The latter decay was 
extremely difficult to observe in CLEO II/II.V, which did not have 
a RICH detector. (Right)
The penguin dominated decay $B \to K\pi$. 
This 
mode is observed in CLEO III with a branching ratios consistent 
with that found in CLEO II/II.V.} 
\label{fig:rich}
\end{center}
\end{figure*}

\subsection {The CLEO-III Detector : Performance, Modifications and issues}

The CLEO III detector, shown in Figure~\ref{fig:cleo3_det}, consists of 
a new silicon tracker, a new drift chamber, 
and a Ring Imaging Cherenkov Counter (RICH), together  with the 
CLEO II/II.V magnet, electromagnetic calorimeter and muon chambers.
The upgraded detector was installed and commissioned during the fall of 1999 
and spring of 2000. 
Subsequently, operation has been very reliable (see below for a caveat) 
and a very high quality data set has been obtained. 
To give an idea of the power of the CLEO III detector, 
Figure~\ref{fig:rich} (left plot) shows the beam 
constrained mass for the Cabibbo allowed decay $B\to D\pi$ and 
the Cabibbo suppressed decay $B\to DK$ with and without RICH information. 
The latter decay was extremely difficult to observe in 
CLEO II/II.V, which did not have a RICH detector. 
In the right plot of Figure~\ref{fig:rich}, 
the penguin dominated decay $B \to K\pi$ and the tree dominated 
decay $B\to \pi\pi$ are shown. 
Both of these modes are observed in CLEO III with branching ratios 
consistent with those found in CLEO 
II/II.V. and are also in agreement with recent Belle and \babar\ results.
Figure~\ref{fig:rich} is a demonstration that CLEO III performs very well 
indeed.

%
Unfortunately, there is one detector subsystem that is not performing well. 
The CLEO III silicon detector, Si3, has experienced an unexpected loss of efficiency
which is increasing with time. The cause of the inefficiency is unknown. 
The situation is under 
constant evaluation but it is likely that Si3  
will be replaced with a wire vertex chamber for CLEO-c. 
We note that if one was to design a charm factory detector from 
scratch the tracking would be entirely gas based to ensure 
that the detector material was kept to a minimum. 
CLEO-c simulations indicate that a simple six layer stereo tracker 
inserted into the CLEO III 
drift chamber  as a silicon replacement would provide a system 
with superior momentum resolution 
to the current CLEO III tracking system. 
The CLEO collaboration therefore proposes to build such 
a device for CLEO-c at a cost of order \$100,000.

Due to machine issues, CLEO also plans to lower the solenoid field 
strength to 1 T from 1.5 T. The other 
parts of the detector do not require modification. 
The dE/dx and Ring Imaging Cerenkov counters 
are expected to work well over the CLEO-c momentum range. 
The electromagnetic calorimeter 
works well and has fewer photons to deal with at 3-5 GeV than at 10 GeV.
Triggers will work as before. 
Minor upgrades may be required of the Data Acquisition system to 
handle peak data transfer rates.
CESR conversion to CESR-c requires 18 m of wiggler magnets 
at a cost of $\sim$ \$4M and is discussed elsewhere.
The conclusion is that, with the addition of the
replacement wire chamber, CLEO is expected to work well in the 3-5 GeV 
energy  range at the expected rates.

\subsection {Examples of analyses with CLEO-c}

The main targets for the CKM physics program at CLEO-c are 
absolute branching ratio 
measurements of hadronic, leptonic, and semileptonic decays.
The first of these provides an absolute 
scale for all charm and hence all beauty decays. 
The second measures decay constants and the third 
measures form factors and, in combination with theory, allows 
the determination of $V_{cd}$ and $V_{cs}$.

\subsubsection {Absolute branching ratios}

\begin{figure}
\begin{center}
\epsfig{figure=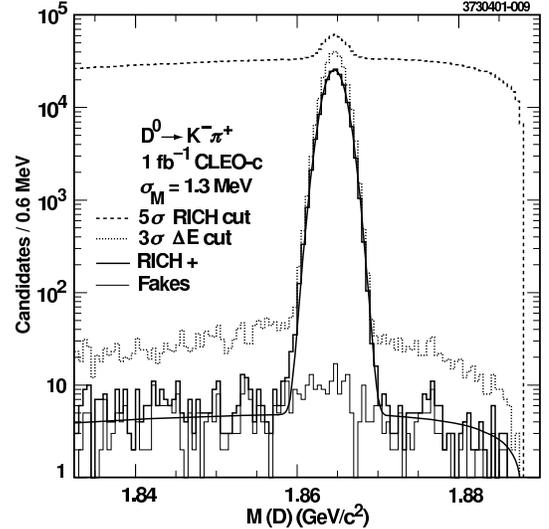,width=7cm,height=7cm}
\caption 
{$K\pi$ invariant mass in $\psi(3770)\to D\bar{D}$ events, 
showing a strikingly clean signal for 
$D\to K\pi$. The y axis is  logarithmic. The S/N $\sim$ 5000/1.}
\label{fig:d0kpi}
\end{center}
\end{figure}

\begin{figure}
\begin{center}
\epsfig{figure=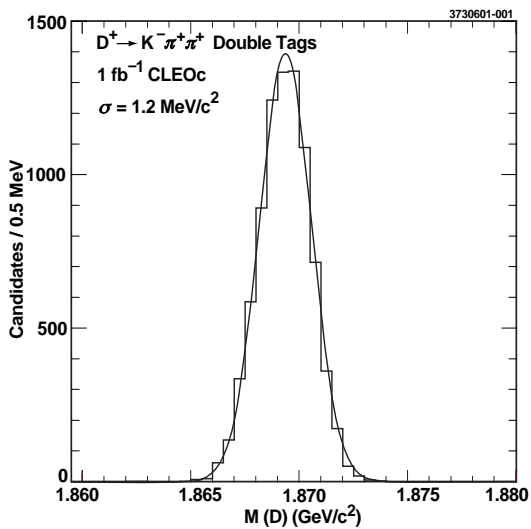,width=7cm,height=7cm}
\caption 
{$K\pi\pi$ invariant mass in $\psi(3770)\to D\bar{D}$  
events, where the other D in the event has 
already been reconstructed. 
A clean signal for $D\to K\pi\pi$ is observed and the absolute 
branching ratio $Br(D\to K\pi\pi)$ is measured by counting events 
in the peak.}
\label{fig:br_dkpipi}
\end{center}
\end{figure}

The key idea is to reconstruct a $D$ meson in as many hadronic modes
as possible. 
This, then, constitutes the tag. Figure~\ref{fig:d0kpi} 
shows tags in the mode $D\to K\pi$. Note the y axis is a log scale. 
Tag modes are very clean. The signal to background ratio is $\sim$ 5000/1 
for the example shown. 
Since $\psi(3770) \to D\bar{D}$, reconstruction of a second $D$ meson in a 
tagged event to a final state X, corrected by the efficiency which is very 
well known, and divided by 
the number of $D$ tags , also very well known, is a measure of 
the absolute branching ratio 
$Br(D\to X)$. Figure~\ref{fig:br_dkpipi} shows the $K^{-}\pi^{+}\pi^{+}$
signal from doubly tagged events. It is essentially background free. 
The simplicity of $\psi(3770) \to D\bar{D}$
events combined with the absence of background allows the determination of 
absolute branching ratios with extremely small systematic errors. 
This is a key advantage of running at threshold.

\subsubsection {Leptonic decay $D_s\to\mu\nu$}

This is a crucial measurement because it provides information which
can be used to extract the weak decay constant, $f_{D_{s}}$. The
constraints provided by running at threshold are critical to extracting
the signal.

The analysis procedure is as follows: 
\begin{enumerate}
\item Fully reconstruct one $D_{s}$;  
\item Require one additional charged track and no additional photons; 
\item Compute the missing mass squared (MM2),  which  peaks at zero for 
a decay where only a neutrino is unobserved. 
\end{enumerate}

\begin{figure*}
\begin{center}
\epsfig{figure=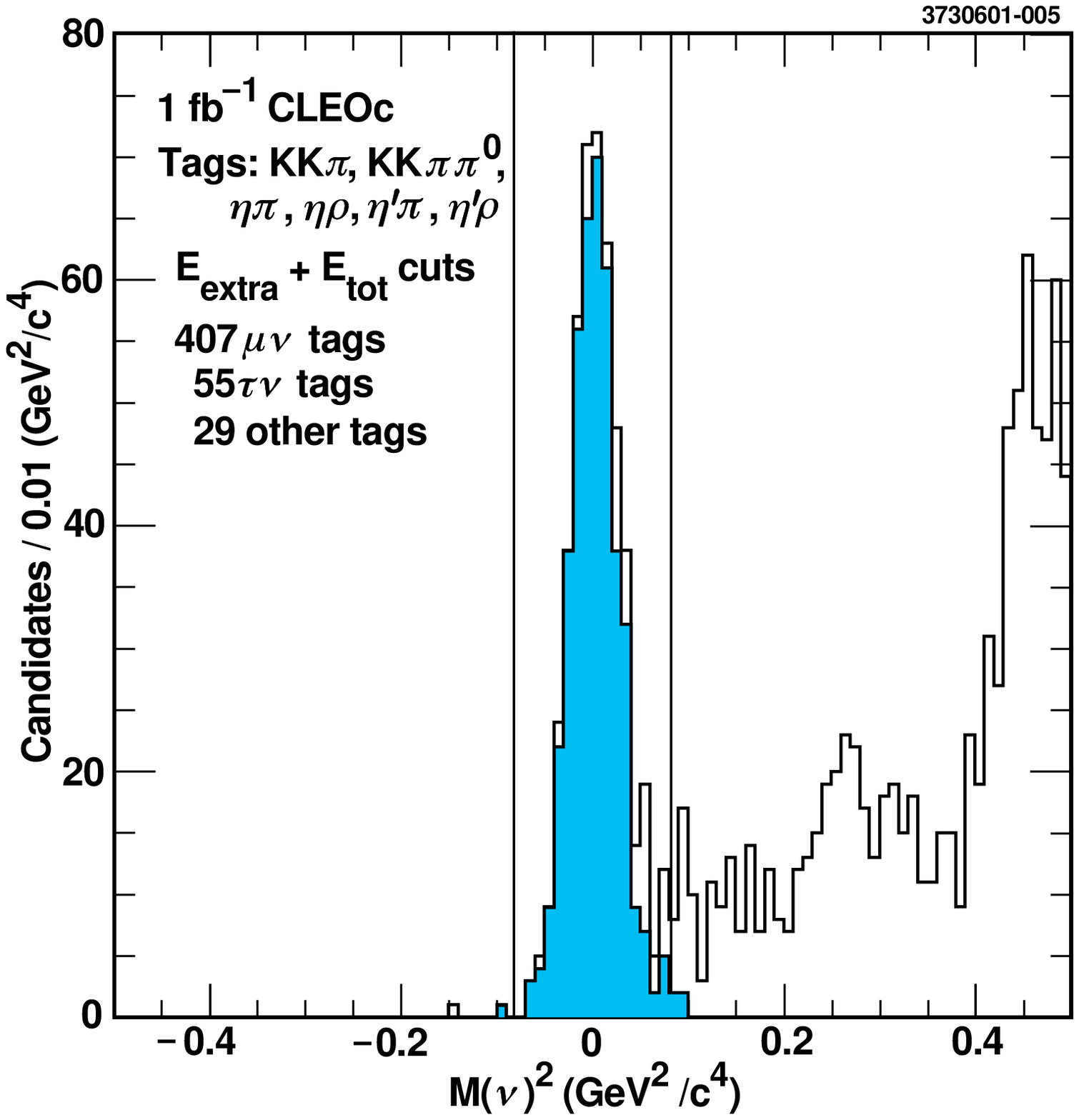,width=6cm,height=6cm}
\epsfig{figure=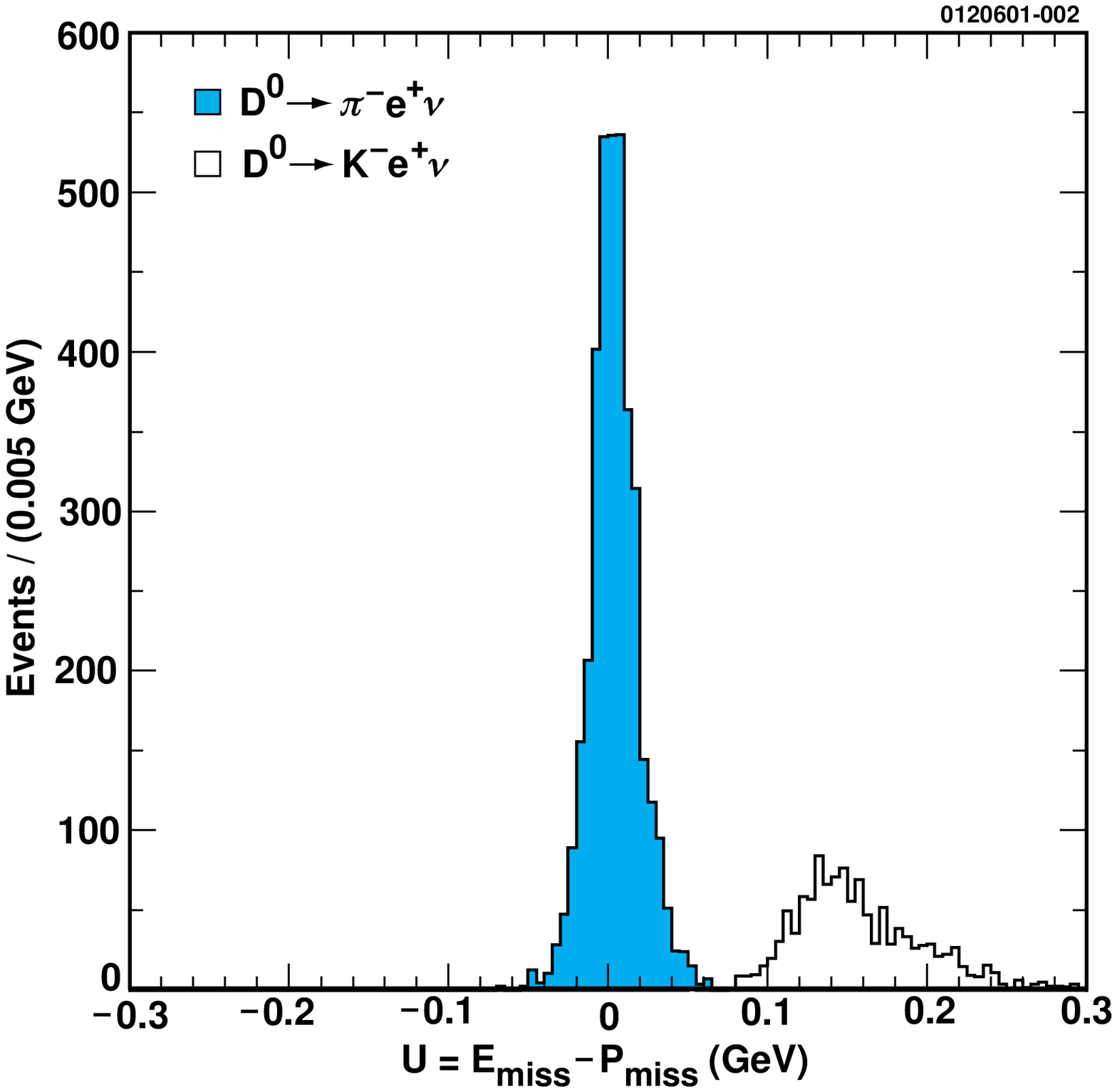,width=6cm,height=6cm}
\caption 
{(Left) Missing mass for $D_{s}D_{s}$ tagged pairs produced 
at $\sqrt{s}=4100$ MeV. Events due to the decay $D_s\to\mu\nu$ are shaded. 
(Right) The difference between the missing energy and missing 
momentum in $\psi(3770)$ tagged events for  
the Cabibbo suppressed decay $D\to \pi \ell\nu$ (shaded). 
The unshaded histogram arises from the ten times more copiously produced 
Cabibbo allowed 
transition $D\to K \ell\nu$, where the K is outside the fiducial volume of 
the RICH.}
\label{fig:munu_pienu}
\end{center}
\end{figure*}

The missing mass resolution, which  is of order $\sim M_{\pi^0}$,
is good enough to reject the backgrounds to this process as shown in
Fig.~\ref{fig:munu_pienu}.
There is no need to identify muons, which helps reduce the systematic error. 
One can inspect the single prong to make sure it is not an electron. 
This provides a check of the background level since the leptonic decay
to an electron is severely helicity-suppressed and no signal
is expected in this mode.
 
\subsubsection {Semileptonic decay $D\to \pi \ell\nu$}

The analysis procedure is as follows:
\begin{enumerate} 
\item Fully reconstruct one D;
\item Select events with one additional electron and one hadronic track; 
\item Calculate the variable $U = E_{miss} - P_{miss}$, which  
peaks at zero for semileptonic decays.
\end{enumerate}

Using the above procedure results in the right-hand plot of 
Figure~\ref{fig:munu_pienu}. 
With CLEO-c, for the first time it will become possible to make 
absolute branching ratio and 
absolute form factor measurements of every charm meson 
semileptonic pseudoscalar to 
pseudoscalar  and pseudoscalar to vector transition. 
This will be a lattice calibration data set without 
equal. Figure~\ref{fig:error_br} graphically shows the improvement 
in absolute semileptonic branching ratios that CLEO-c will make.

\subsection {Run Plan}

CLEO-c must run at various center of mass energies in order to
achieve its physics goals. The ``run plan'' currently used
to calculate the physics reach is given below. Note that item 1 is prior 
to machine conversion and the remaining items are post machine conversion.
\begin{enumerate}
\item 2002: $\Upsilon$'s --  1-2 $fb^{-1}$ each at 
$\Upsilon(1S), \Upsilon(2S), \Upsilon(3S)$ \\
Spectroscopy, electromagnetic transition matrix elements, the leptonic width. 
$\Gamma_{ee}$, and searches for the yet to be discovered 
$h_b, \eta_b$ with 10-20 times the existing world's data sample.     
\item 2003: $\psi(3770)$ -- 3 $fb^{-1}$ \\
30 million events, 6 million tagged D decays (310 times MARK III).        
\item 2004: 4100 MeV -- 3 $fb^{-1}$ \\
1.5 million $D_{s}D_{s}$ events, 0.3 million tagged $D_s$ decays 
(480 times MARK III, 130 times BES).        
\item 2005: $J/\psi$ -- 1 $fb^{-1}$ \\
1 Billion $J/\psi$ decays (170 times MARK III, 20 times BES II).
\end{enumerate}

\subsection{Physics Reach of CLEO-c}

\begin{figure}
\begin{center}
\vspace{0.5cm}
\epsfig{figure=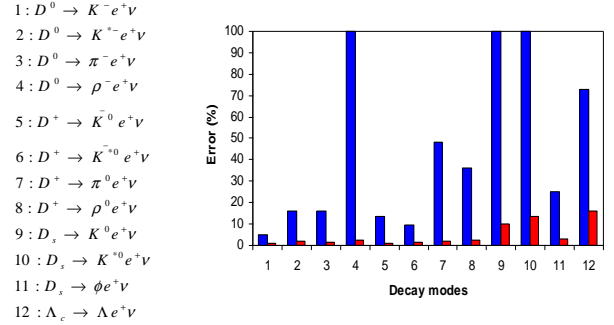,height=8cm,angle=-90}
\caption 
{Absolute branching ratio current precision from the PDG 
(left entry) and precision 
attainable at CLEO-c (right entry ) for twelve semileptonic charm decays.}
\label{fig:error_br}
\end{center}
\end{figure}

\begin{figure}
\begin{center}
\epsfig{figure=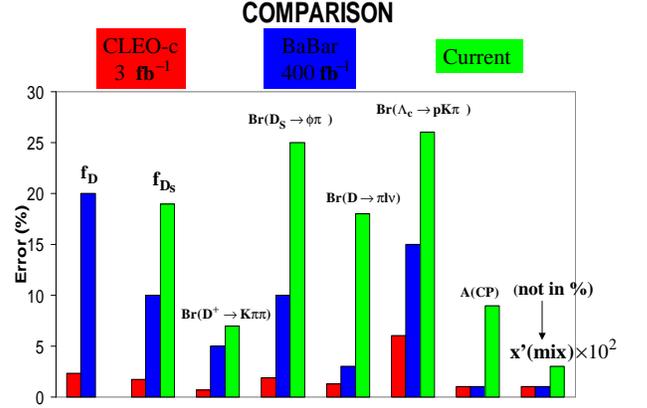,height=8cm,angle=-90}
\caption 
{Comparison of CLEO-c  (left) \babar\ (center) and PDG2001 (right) 
for eight physics quantities indicated in the key.}
\label{fig:comparison}
\end{center}
\end{figure}

Several talks to the E2 working group addressed the competition CLEO-c 
will face from BESII/III~\cite{BESzhao}, 
\babar~\cite{LOU}, and experiments at 
hadron machines ~\cite{FOCUS},\cite{DTEV}.
Tables~\ref{tab:charm}, \ref{tab:ckm} , and \ref{tab:comparison}, 
and Figures~\ref{fig:error_br} and 
\ref{fig:comparison} summarize the CLEO-c measurements 
of charm weak 
decays, and compare the precision obtainable with CLEO-c 
to the expected precision at \babar,  
which expects to have recorded 500 million charm pairs by 2005. 
CLEO-c clearly achieves far greater 
precision for many measurements. 
The reason for this is the ability to measure absolute branching 
ratios by tagging and the absence of background at threshold. 
In those topics where CLEO-c is 
not dominant, it remains comparable or complementary to the \bm-factories. 

Also shown in Table~\ref{tab:comparison} is a summary of the data set size 
for CLEO-c and BES II at the $J/\psi$ and $\psi'$, 
and the precision with which R, the ratio of the $e^{+}e^{-}$ annihilation 
cross section into hadrons to $\mu$ pairs, can be measured. 
Since the CLEO-c data sets are over an order of magnitude larger, the 
precision with which R is measured is a factor of three higher. 
In addition, the CLEO detector is vastly superior to the BES II detector. 
Taken together, the CLEO-c datasets at the $J/\psi$ and $\psi'$ will be 
qualitatively and quantitatively superior to any previous dataset 
in the charmonium sector thereby 
providing discovery potential for glueballs and exotics without equal.

\begin{table*}[t]
\begin{center}
\caption[]
{Summary of CLEO-c charm decay measurements.}
\label{tab:charm}
\begin{tabular}{c|c|c|c|c|c}
Topic & Reaction & Energy & $L $        & current     & CLEO-c \\ 
      &          & (MeV)  & $(fb^{-1})$ & sensitivity & sensitivity \\ 
\hline
\multicolumn{1}{c}{Decay constant} &
\multicolumn{5}{c}{}\\
\hline
$f_D$ & $D^+\to\mu^+\nu$ & 3770 & 3 & UL & 2.3\% \\
\hline
$f_{D_s}$ & $D_{s}^+\to\mu^+\nu$ & 4140 & 3 & 14\% & 1.9\% \\
\hline
$f_{D_s}$ & $D_{s}^+\to\mu^+\nu$ & 4140 & 3 & 33\% & 1.6\% \\
\hline
\multicolumn{2}{c}{Absolute Branching Fractions} &
\multicolumn{4}{c}{}\\
\hline
\multicolumn{2}{c|}{$Br(D^0 \to K\pi)$} & 3770 & 3 & 2.4\% & 0.6\% \\
\hline
\multicolumn{2}{c|}{$Br(D^+ \to K\pi\pi)$} & 3770 & 3 & 7.2\% & 0.7\% \\
\hline
\multicolumn{2}{c|}{$Br(D_s^+\to\phi\pi)$} & 4140 & 3 & 25\% & 1.9\% \\
\hline
\multicolumn{2}{c|}{$Br(\Lambda_c\to pK\pi)$} & 4600 & 1 & 26\% & 4\% \\
\end{tabular}
\end{center}
\end{table*}

\begin{table}[t]
\begin{center}
\caption[]
{Summary of direct CKM reach with CLEO-c}
\label{tab:ckm}
\begin{tabular}{c|c|c|c|c|c}
Topic & Reaction & Energy & $L $        & current     & CLEO-c \\ 
      &          & (MeV)  & $(fb^{-1})$ & sensitivity & sensitivity \\ 
\hline
$V_{cs}$ & $D^0\to K\ell^+\nu$ & 3770 & 3 & 16\% & 1.6\% \\
\hline
$V_{cd}$ & $D^0\to\pi\ell^+\nu$ & 3770 & 3 & 7\% & 1.7\% \\
\end{tabular}
\end{center}
\end{table}

\begin{table*}[t]
\begin{center}
\caption[]
{Comparison of CLEO-c reach to \babar\ and BES}
\label{tab:comparison}
\begin{tabular}{c|c|c||c|c|c}
Quantity & CLEO-c & \babar\  & Quantity & CLEO-c & BES-II \\ 
\hline
$f_D$ & 2.3\% & 10-20\%    & \#$J\psi$ & $10^9$ & $5\times 10^7$ \\ 
\hline
$f_{D_s}$ & 1.7\% & 5-10\% & $\psi'$   & $10^8$ & $3.9\times 10^6$ \\
\hline
$Br(D^0 \to K\pi)$ & 0.7\% & 2-3\% & 4.14 GeV & $1\,fb^{-1}$ &$23\,pb^{-1}$\\ 
\hline
$Br(D^+ \to K\pi\pi)$ & 1.9\% & 3-5\%  & 3-5 R Scan & 2\%  & 6.6\% \\
\hline
$Br(D_s^+\to\phi\pi)$ & 1.3\% & 5-10\% & \multicolumn{3}{c}{} \\
\end{tabular}
\end{center}
\end{table*}

\subsection{CLEO-c and Future Competition}

BES/BEPC is currently proposing to upgrade the machine and 
detector~\cite{BESzhao}.
In response to the CESR-c/CLEO-c proposal, the design goal for 
the machine, BEPC II, was recently changed from a peak luminosity of
$ 5 \times 10^{31} {\rm cm}^{-2}{\rm s}^{-1} $ to a two
ring machine with a peak luminosity in excess of 
$ 10^{33} {\rm cm}^{-2}{\rm s}^{-1} $. A completely new detector, BES III,
would be built possibly around an electromagnetic calorimeter
made of BGO crystals from the L3 experiment. The detector design is evolving 
and is the subject of a planned workshop in Beijing in October 2001.
As now envisaged BEPCII/BESIII would come on line around 2006
and would accumulate a data sample
one order of magnitude larger than CLEO-c. 
The physics program  of BES III is identical to CLEO-c. 
For BES III to make a significant impact it 
is absolutely essential that the detector be as good as the CLEO-c detector.
If that can be achieved, the significantly larger luminosity of BEPCII over 
CESR-c is likely to be a considerable advantage for new physics reach.
For CKM physics, theory will have to sharpen for the larger statistics of 
BES III to be used to full advantage. 

A program is underway at TJNAL to systematically explore the light mesons
with masses up to 2.5 GeV/c$^{2}$ using photoproduction 
with the high quality low emittance CW photon beams available there. 
The program will be capable of exploring both
light meson states and searching for exotic states in this mass region. A new detector 
is proposed along with an upgrade of CEBAF to 12 GeV. The target date for completion 
of construction is 2006. The goals of HALL-D and CLEO-c have some overlap
but there is also complementarity. CLEO-c is focusing on glue rich states and vector
hybrids both light and heavy. Hall-D is focused on states with exotic quantum numbers.

There is a proposal from the GSI accelerator in Germany for a High Energy Storage Ring
(HESR) for antiprotons. One part of the program of this facility  
will be a search for gluonic excitations, glueballs and hybrids in the charmonium sector.
This interesting facilty was not discussed in the E2 group as GSI was not represented.
However, charmonium studies are likely to be complementary to CLEO-c.

\subsection{CLEO-c Physics Impact} 

CLEO-c will provide crucial validation of Lattice QCD, 
which  will be able to  calculate many quantities with claimed
accuracies of 1-2\%. 
The CLEO-c decay constant and semileptonic data will provide a ``golden'', 
and timely test 
while CLEO-c QCD and charmonium data will provide additional benchmarks.

CLEO-c will provide, in a timely fashion,  dramatically improved  knowledge 
of absolute charm branching fractions, 
which are now contributing significant errors to measurements involving 
b's. CLEO-c will significantly improve knowledge of those CKM matrix elements 
which are now not very well known. In particular, 
$V_{cd}$ and $V_{cs}$ will be determined directly by CLEO-c data and LQCD,
or other theoretical techniques. 
$V_{cb}, V_{ub}, V_{td}$ 
and $V_{ts}$ will be determined with enormously improved precision 
using \bm-factory data and lattice gauge results once the 
CLEO-c program of lattice validation is complete. 
Table~\ref{tab:ckm_summary} gives a summary of the situation.
CLEO-c data alone will also allow new tests of the unitarity of the CKM matrix.
The unitarity of the second 
row of the CKM matrix will be probed at the 3\% level, which 
is comparable to our current knowledge of the first row.
CLEO-c data will also test unitarity by measuring 
the ratio of the long sides of the 
squashed  $cu$ triangle to 1.3\%.

Finally the potential to observe new forms of matter, glueballs, 
hybrids, etc in $J/\psi$ decays, and new 
physics through sensitivity to charm mixing, {\it CP} violation, 
and rare decays provides a discovery 
component to the program.

\begin{table}[t]
\begin{center}
\caption[]
{Current knowledge of CKM matrix elements (row one). Knowledge of 
CKM matrix elements after CLEO-c (row two). 
See the text for further details.}
\label{tab:ckm_summary}
\begin{tabular}{c|c|c|c|c|c}
$V_{cd}$ & $V_{cs}$ & $V_{cb}$ & $V_{ub}$ & $V_{td}$ & $V_{ts}$ \\
\hline
7\% & 16\% & 5\% & 25\% & 36\% & 39\% \\
\hline
1.7\% & 1.6\% & 3\% & 5\% & 5\% & 5\% \\
\end{tabular}
\end{center}
\end{table}

\section{$\lowercase{e^{+}e^{-}}$ \bm-Factories and 
Their Plans for the Future}

The two asymmetric \bm-factories, PEP-II and KEKB, have achieved reliable
operation at high luminosities of a few $10^{33}{\rm cm}^{-2} {\rm s}^{-1}$ in
a remarkably short period of time after their startup. These luminosities
have enabled their experiments, \babar\ and Belle, respectively, to
observe {\it CP} violation in the decays of the $B^{o}$ meson. Operational
experience with both machines has now led to plans for incremental
upgrades which eventually are expected to produce luminosities of
$10^{35}{\rm cm}^{-2}{\rm s}^{-1}$. For the purposes of this report, we will
refer to these as ``super \bm-factories'', with a lower case `s'. While
this is happening, hadron collider experiments at the Tevatron, 
CDF and D0, will begin to produce \bm\ physics results that will compete
with, and in some cases exceed, the sensitivity of the $e^{+}e^{-}$
\bm-factories. Dedicated experiments at the Tevatron and the LHC,
BTeV, and LHCb, and the two large general
purpose experiments at the LHC, CMS and ATLAS, will begin to contribute
at very high levels of sensitivity to the study of {\it CP} violation
and rare decays in the $B$ system, starting around 2007. The SLAC
group has proposed a response to this, which we refer to as the
``Super \bm-factory'', which has a luminosity
goal of  $10^{36}{\rm cm}^{-2}{\rm s}^{-1}$. We write this with an uppercase
`S' to emphasize that it is aiming at a factor of 10 higher luminosity
than superKEK.
This requires a new machine and
a very significant upgrade of the \babar\ detector. KEK seems, at present, to
have no plans to pursue \bm\ physics after the dedicated hadron collider $B$ 
experiments appear on the scene. We present the plans for the two phases of
\bm-factory upgrade, emphasizing physics reach, and compare their reach to the
physics reach of the hadron collider experiments that will be coming on
in the same period. This part of the report is based on 
the following set of talks to the E2 working group ~\cite{Hitlin} - ~\cite{Robertson}, 
much lively discussion
and much work during the summer study, especially by the E2 subgroup
on Super \bm-factories organized by David Hitlin~\cite{SuperBaBar}. 
The projected evolution of luminosity in these
machines is shown in Table~\ref{tab:lumin_evol}.

In addition to these two circular machines, there are proposals to
construct multi-hundred GeV center of mass energy  $e^{+}e^{-}$
Linear Colliders. This has raised the prospect of further running
on the $Z$-pole, where: the $b$-quark cross section is very high, $\sim$7nb;
where all species of $B$ mesons and baryons are produced; 
there is significant boost for time-dependent studies; and  
the events are quite clean allowing flavor tagging to be done efficiently.
While most of the time the machine will operate at a center of mass
energy well above the $Z$-pole, it is possible to invent a scheme where
continuous $Z$-pole running is possible. Such a scheme is proposed for
TESLA at
DESY where there is a second beam for a Free Electron Laser. Pulses can
be stolen from that to form a so-called Giga-$Z$ machine. The physics reach  of
this machine is explored, some areas in which it can do unique studies
are described, and its sensitivities are compared with those of the circular 
$e^{+}e^{-}$ machines and the hadron colliders.

\begin{table}[t]
\begin{center}
\caption[]
{Predicted Evolution of Luminosity and Number of Produced $B$'s in Asymmetric
$B$ Factories}
\label{tab:lumin_evol}
\begin{tabular}{c|c|c|c|c|c|c} \hline
               & KEKB    & KEKB   & PEPII   & PEPII & super & Super \\
               & 2001    & 2005   & 2001    & 2005  & KEKB  & \babar \\ 
               &         &        &         &       & $>$ 2007
                                                    & $>$ 200X  \\ \hline
$L\times 10^{33}$ & 4.1  & 10     & 3       & 10    & 100   & 1000  \\ \hline
$B$'s/$10^{7}$s& 8.2$\times\,10^{7}$  & 2$\times\,10^{8}$ & 
6$\times\, 10^{7}$ &  2$\times\,10^{8}$ & 2$\times\,10^{9}$ &
2$\times\,10^{10}$ \\ \hline
\end{tabular}
\end{center}
\end{table}

\subsection{KEKB/Belle Upgrade plans}

KEK plans for call for an upgrade to $10^{35}{\rm cm}^{-2}{\rm s}^{-1}$, 
which corresponds to
$10^{9}$ \bm\ pairs per year. Towards the end of this period, which they
see as extending to around 2007/8, they expect to be overtaken by competition 
from hadron collliders. However, they believe that they will have significant
advantages with respect to hadron colliders in terms of
\begin{itemize}
\item $\pi^{o}$ and $\gamma$ detection efficiency, and
\item smaller backgrounds.
\end{itemize}
They look to techniques such as greater reliance on vertex separation cuts 
and full reconstruction tagging to reduce backgrounds below what they are 
today. With the improved backgrounds
obtained with a detachment cut of about 2$\sigma$, 
they believe it will be possible to study decays with branching
fractions at the level of 5$\times 10^{-7}$.
Examples of decays that would then be accessible 
are $B^{-}\rightarrow K^{*o}K^{-}$
and decays such as $B^{+}\rightarrow D^{+}K_{s}$ and 
$B^{+}\rightarrow  D^{o}K^{+}$, which can be used to measure the CKM angle
$\gamma$.
In full reconstruction tagging, as many $B$'s as possible are 
fully reconstructed
and then one studies the remnants, which must all be from the other $B$. This 
technique helps especially with states containing neutrinos, such
as 
\begin{eqnarray}
b & \rightarrow & ul\nu \\
B & \rightarrow & \mu \nu \\
b & \rightarrow & s \nu \bar{\nu} 
\end{eqnarray}
The technique relies on the detector's hermeticity.

The conclusion is that there are many significant physics studies they can
do with approximately 5 years of running at a  luminosity of $10^{35}$. 
The machine upgrade is an extrapolation of the current KEK configuration.
It was discussed in section M2~\cite{Ohnishi}.

Operation at $10^{35}$ has implications for the detector and the IR. The
rates from collisions will be significantly higher which will lead to
larger occupancy. Trigger rates and rates through the data acquisition system 
will be higher. There will be more synchrotron radiation, which
will have to be removed by masking. There may be larger vacuum pressure
resulting in higher background rates from Touschek scattering. There may
need to be a larger crossing angle which may make it harder to shield
backgrounds efficiently. The final quads may be moved closer to the IP
to reduce $\beta^{*}$. And finally, the background at injection might be
significantly worse.

It is planned to use a 1 cm radius beampipe.
Particle backgrounds will be controlled by massive masks around the inner
vertex detectors, on the upstream beampipes and at other ``weak spots''.
Nevertheless, the first few layers of the silicon vertex detector
will have high occupancy and will be replaced by pixel detectors.
Beampipe heating due especially to Higher Order Modes (HOM) requires that
the beam pipe be water cooled. The Central Drift Chamber is undergoing a 
modification in 2002 to replace the two inner layers with a small cell 
chamber. It is expected to be able to handle superKEK rates. The CsI(Tl) 
calorimeter is slow and something may need to be done to it. The RPCs
in the muon system already suffer from inefficiency due to local
deadtime and will probably need to be replaced with wire chambers.
The data acquisition system will also have to be upgraded.

The upgrade to $10^{35}$ is believed to be feasible from a machine point of 
view. The detector will need several upgrades but these appear feasible 
as well. 
The physics case is based on the cleanliness of the signals and the ability
to study modes that are very hard to measure in hadron colliders -- modes
which include $\pi^{o}$'s and $\nu$'s. After several years of running
at  $10^{35}$, the $B$ physics program at KEK will probably end. A further
push in luminosity would require a new machine configuration and 
a new detector and is not in their current plans.

\subsection{PEP-II/\babar\ Upgrade Plans: Super B Factory and Super\babar}

PEP-II and \babar\ expect to achieve an integrated luminosity of 
500 fb$^{-1}$ (0.5 ab$^{-1}$)  by around 2005.  With that, they expect
to achieve the following errors on the unitarity angles $\beta$ and $\alpha$: 
\begin{eqnarray}
\sin 2\beta & \approx & 0.04 \\
\sin 2 \alpha & \approx & 0.14 
\end{eqnarray}
For details of these estimates and a discussion of the prospects and complications
in the measurement of $\alpha$ and $\gamma$ see ~\cite{SuperBaBar}.
Although the combined \babar\ and Belle integrated luminosity will be 
about 1 ab$^{-1}$ at this
point and PEP-II will be delivering about 0.2 ab$^{-1}$/year,
a new generation of hadron collider experiments will be positioned to
dominate the study of {\it CP} violation and rare and Standard Model
forbidden processes in  $B$ decays. A recent study has outlined a possible
path for achieving a luminosity of $10^{36}{\rm cm}^{-2}{\rm s}^{-1}$ 
in $e^{+}e^{-}$ collisions.
This corresponds to 10 ab$^{-1}$/year
and requires a new machine configuration and a very substantial
upgrade of the \babar\ detector, which involves complete replacement or
major revision of many components. The goal is to be competitive with 
the next generation hadron collider experiments, at least in the area
of $B_{d}$ and $B_{u}$ physics. Because of the experimental
constraints of threshold production and the low backgrounds in 
$e^{+}e^{-}$ physics, certain measurements could be made with this
facility that might not be possible to do at hadron colliders.

Details of the new machine can be found in the M2 summary elsewhere in these proceedings. 
The machine could be located
either in the PEP tunnel, where it would replace PEP-II, or in the tunnel
for the SLC arcs.
If located in the PEP tunnel, PEP-II operation would have to stop
for about 1 year while the new machine components were installed.

\subsubsection{Physics Case for 10 {\rm ab}$^{-1}$/yr  $e^{+}e^{-}$ Facility}

There are a variety of interesting topics which can be addressed at such
a facility. These include both precision tests of the consistency of the
Standard Model predictions and discovery of, or sorting out of, new phenomena
beyond the Standard Model. A list of interesting processes are:
\begin{itemize}
\item Improvement in {\it CP} asymmetry measurements
\begin{itemize}
\item $\sigma(\sin 2\beta)\approx 0.01$ for {\it J}/$\psi\,K_{s}$
\item $\sin 2\beta$ will be measured with good precision in many modes which 
     provides an important consistency check
\item $\sin 2\alpha$($A_{{\it CP}}$) and $\sin \gamma$ can be measured 
\end{itemize}
\item Measurement of some particularly challenging 
two and three body branching fractions, for example $B^{o}\rightarrow \pi^{o}\pi^{o}$
\item Measurement of $f_{B}$ 
to useful precision to check lattice predictions
\item Interesting sensitivity to rare $B$, $D$, and $\tau$ decays, such as 
$\tau \rightarrow \mu \gamma$
\item High precision measurements of semileptonic decay distributions,
especially the precision measurement of $V_{ub}$.
\end{itemize}
These topics were studied in the context of a high luminosity next 
generation $e^+e^-$ B Factory at the ``Beyond $10^{34}$ Workshop"~\cite{10E34}
in Michigan during June 2000, at the follow-up session at the Fourth International 
Conference on B physics and {\it CP} Violation in Ise Shima, Japan in February 2001~\cite{Tony} and 
were the focus of an E2 subgroup at Snowmass~\cite{SuperBaBar}.

\subsubsection{Experimental Considerations}

Both the rates from the beam collisions and from backgrounds will be much 
higher than present. In particular, the overall loss rates will be 
about 1000 times
the present rates. The beam lifetime will be only around 10 minutes so the
machine will be filled continuously during the store. At a luminosity of
$10^{36}{\rm cm}^{-2}{\rm s}^{-1}$, there are
\begin{itemize}
\item 50 kHz of Bhabha scatters,
\item $\sim$7 kHz of other physics events, and
\item $O$($\sim$10kHz) of triggerable machine associated background in the
detector acceptance.
\end{itemize}

\subsubsection{Detector Issues}

\begin{figure*}[t]
\begin{center}
\epsfig{figure=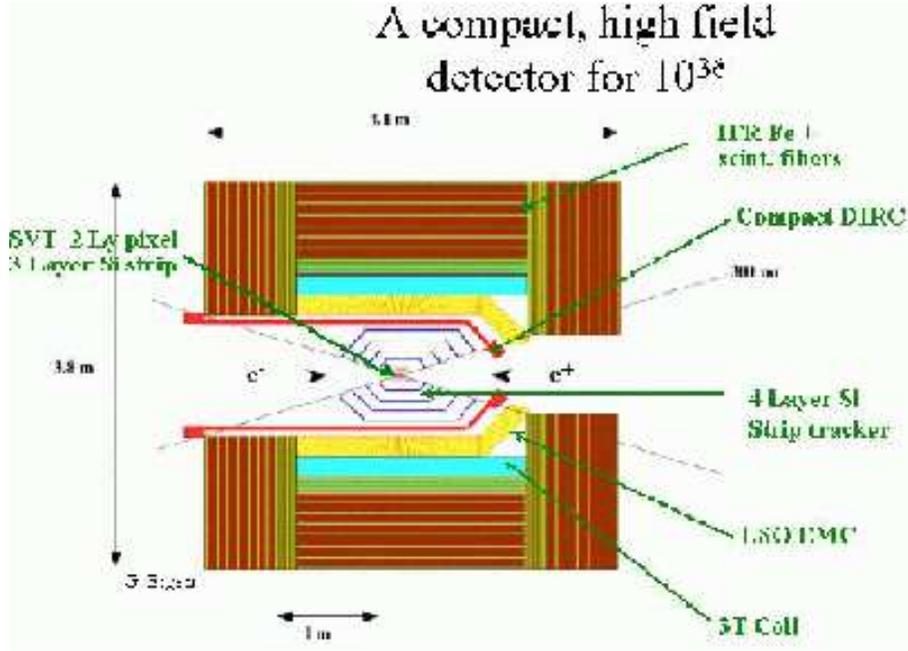,width=12cm}
\caption {Schematic of a Compact Detector Design for Super\babar}
\label{fig:compact}
\end{center}
\end{figure*}

Most of the \babar\ subsystems will have to undergo some modification or
replacement to handle the much higher rates of the new machine.
To carry out the program, the overall performance, in terms of resolution,
efficiency, and background rejection,  must be similar to that of \babar.
The detector must retain its high degree of hermeticity as well. 
Table~\ref{tab:superb_det_mods} summarizes the problems that affect current
\babar\ detectors at these high luminosities and indicate possible solutions.
One concept for the replacement
detector, a very compact detector based on a high field solenoid,
is shown in Fig.~\ref{fig:compact}. The solenoid has a radius of about
0.75m and a field of 3 Tesla. The central vertex detector consists
of two layers of pixel detector and a three layer silicon strip detector.
The central drift chamber is replaced by a 4 layer silicon strip tracker,
which is much more compact. The combination of the high field and the
high precision tracking permit the detector to achieve momentum resolution
comparable to \babar. The expensive electromagnetic crystal calorimeter 
has a small radius, which lowers the cost.

In addition to detector modifications, a faster and more selective trigger
and a higher speed, higher capacity Data Acquisition system must be 
implemented. While difficult compared to the existing \babar\ experiment,
the triggering and data acquisition problem is far less of a challenge
than must be met at the Tevatron or LHC so this is not considered
an insurmountable task. Data analysis will benefit from the projected
continued drop in cost of computing cycles and data storage.

There are substantial uncertainties in the detector requirements
due to the difficulty in estimating the various backgrounds. It is clearly
important to implement a realistic machine lattice and IR design 
to provide predictions for the very large backgrounds that will exist at
Super\babar, especially backgrounds due to continuous injection. These studies
were foreseen, but had not been performed at the time of Snowmass.

There are many questions about the cost and availability of suitable 
detector technologies which will need to be studied before the detector design can be
finalized. We give four examples. (1) To maintain the vertex resolution of 
\babar\ and 
withstand the radiation environment, pixels with a material budget of 0.3\% $X_o$ per layer
are proposed. 
Traditional pixel detectors
which consist of a silicon pixel array bump-bonded to a readout chip 
are at least 1.0\% $X_o$. To obtain less material, monolithic 
pixel detectors are suggested.
This technology has never been used 
in a particle physics experiment. (2) As a drift chamber cannot cope with the large rates and
large accumulated charge, a silicon microstrip tracker has been proposed. At these low energies
track parameter resolution is dominated by multiple Coulomb scattering. Silicon microstrip
technology is well tested but is usually used at this energy for vertexing, not tracking.
Realistic simulations need to be performed to establish if
momentum resolution as good as \babar\ can be achieved with the 
large amount of material present in the silicon tracker.
If not, we suggest a TPC, possibly readout with a Gas Electron Multiplier, or MICROMEGAS,  
be explored as an alternative to the silicon tracker  (3) There is no established crystal technology 
to replace the Csi(Tl). There are some candidate materials (see the 
Super\babar\
document for details) but the most attractive have not been used in
a calorimeter previously. (4) There is no known technology for the light 
sensor for the SuperDIRC.

\begin{table}[t]
\begin{center}
\caption[]
{Modifications to the \babar\ detector for Super\babar.}
\label{tab:superb_det_mods}
\begin{tabular}{l|l|l} \hline
\babar  & Super\babar & Reason for change  \\
Detector & Detector &                     \\ \hline
Silicon Strips & Silicon Pixels & Occupancy \\ \hline
Drift Chamber  & Silicon Tracker & accumulated charge \\ 
               & or TPC          &                 \\ \hline
DIRC           & super DIRC      & Remove water standoff box due   \\
               &                 & to high background Cerenkov \\
               &                 & light and replace with new \\ 
               &                 & optics   \\ \hline
ECAL CsI(Tl)   & new rad hard,   & CsI(Tl) has a long decay  \\ \hline
               & crystal         & time and is not rad hard  \\   
IFR(RPCs)      & scintillators   & Occupancy \\ \hline
\end{tabular}
\end{center}
\end{table}

\subsubsection{Comparison with Hadron Collider Experiments}

Since the goal of the Super \bm-Factory and Super\babar\ upgrades are to
enable the $e^{+}e^{-}$ machine to compete with future hadron collider
experiments, it is important to make a realistic evaluation of the 
sensitivities of all these experiments over a wide range of final states.
Such projections are, of course, somewhat uncertain. 
%
The sensitivities of future hadron collider experiments have been determined from
detailed and sophisticated simulations of signals and backgrounds. As these simulations
are an approximation to reality, the performance of LHCb and BTeV may be
somewhat better or somewhat worse than the simulations predict.
Projections for Super\babar\ are, at this point, 
mainly done by scaling from \babar\ experience assuming that the new
detector, which still has many open R\&D issues, will achieve the same 
efficiency that \babar\ now achieves even though the luminosity will be a 
factor of 300 higher. More
realistic studies need to be performed before a full comparsiion 
between Super\babar\ and the hadron collider experiments is made.

\begin{table}
\begin{center}
\caption[]
{Comparison of the number of tagged $B^{o}\rightarrow \pi^{+}\pi^{-}$
 in Super\babar\ and BTeV}
\label{tab:comp_pi_pi}
\begin{tabular}{l|c|c|c|c|c|c|c} \hline
       & $L(\rm{cm}^{-2}\rm{s}^{-1})$ & $\sigma$ & $B^{o}$/10$^{7}$s &
$\epsilon$ & S/B & $\epsilon D^{2}$ & tagged  \\ \hline
$e^{+}e^{-}$ & 10$^{36}$ & 1.1 nb & 1.1$\times$10$^{10}$ & 0.3   & 0.7 & 0.3 &
3600 \\ \hline
BTeV & 2$\times$10$^{32}$ & 100$\mu$b & 1.5$\times$10$^{11}$ & 0.037 & 3.0 &
 0.1 & 2370 \\ \hline
\end{tabular}
\end{center}
\end{table}

\begin{table}
\begin{center}
\caption[]
{Comparison of the number of tagged $B^{+}\rightarrow D^{o}K^{+},
D^{o}\rightarrow  K^{+}\pi^{-}$ with  in Super\babar\ and BTeV
(product of all branching
factions taken as B=$1.7\times 10^{-7}$) }
\label{tab:comp_B_DK}
\begin{tabular}{l|c|c|c|c|c|c} \hline
       & $L(\rm{cm}^{-2}\rm{s}^{-1})$ & $\sigma$ & $B^{o}$/10$^{7}$s &
$\epsilon$ & S/B & tagged  \\ \hline
$e^{+}e^{-}$ & 10$^{36}$ & 1.1 nb & 1.1$\times$10$^{10}$ & 0.5   &  & 
600 \\ \hline
BTeV & 2$\times$10$^{32}$ & 100$\mu$b & 1.5$\times$10$^{11}$ & 0.014 & 1.0 &
300 \\ \hline
\end{tabular}
\end{center}
\end{table}

\begin{table*}
\begin{center}
\caption[]
{Comparison of {\it CP} Reach of Hadron Collider Experiments and
Super\babar. The last column is a prediction of which kind of facility
will make the dominant contribution to each physics measurement.}
\label{tab:comp_e_had_cp}
\begin{tabular}{|l|c|c|c|c|c|c|} \hline
       & BTeV       &  LHCb     &  \babar   & 10$^{35}$  & 10$^{36}$  &   \\
       &  10$^{7}$s & 10$^{7}$s &  Belle   & 10$^{7}$s  &  10$^{7}$s &   \\
       &            &           & (2005)   &            &            &   \\ 
\hline
$\sin 2\beta$ &  0.011 &  0.02 &  0.037 &  0.026 &  0.008 & Equal \\
$\sin 2\alpha$&  0.05  &  0.05 &  0.14  &  0.1   &  0.032 & Equal \\
$\gamma\,[B_{s}(D_{s}K)]$ & $\sim$7$^{o}$ &   &   &   &   & Had  \\
$\gamma\,[B(DK)]$ & $\sim$2$^{o}$ &   &$\sim$20$^{o}$   &   &1-2.5$^{o}$&Equal \\
$\sin 2\chi$ &  0.023 &  0.04  &  -  &  -  &  - & Had \\
BR($B\rightarrow \pi^{o}\pi^{o}$) & -  &  -  & $\sim$20\% & 14 \% & 6\% &
$e^{+}e^{-}$ \\
$V_{ub}$ & -  &  -  & $\sim$2.3\% & $\sim$1\%  & $\sim$1\% &
$e^{+}e^{-}$ \\
         &    &     &             &  (sys)     & (sys)  &   \\ \hline
\end{tabular}
\end{center}
\end{table*}

\begin{table*}
\begin{center}
\caption[]
{Comparison of Reach of Hadron Collider Experiments and Super\babar 
for Rare Decays of $B_{u}$ and $B_{d}$ Mesons. Entries are either
branching fraction sensitivities, if they have negative exponents,
or signal yields. An $\star$ indicates that the entry below is claimed to be 
the best measurement. The numbers in parentheses in column 1 are the 
branching fractions used in the calculations.}
\label{tab:comp_e_had_rareb}
\begin{tabular}{l||c|c|c||c|c|c} \hline
       & \multicolumn{3}{c||}{Hadronic Exp} &\multicolumn{3}{c}{B-Factory} \\
\hline
Decay Mode   & CDF  &  BTeV     &  ATLAS   & \babar & 10$^{35}$  & 10$^{36}$ \\
             &/D0   & /LHCb     &  /CMS    &/Belle &            &           \\
(Br Ratio)   & (2$fb^{-1}$) &  10$^{7}$s & (1 year) & (0.5$ab^{-1}$) &
(1$ab^{-1}$) & (10$ab^{-1}$) \\ \hline
$B\rightarrow X_{s}\gamma$ &  &  &  &   & $\star$   &  $\star$  \\
(3.29$\pm$0.21$\pm$)0.21)$\times\,10^{-4}$ &   &  &   & 11K  & 22K & 220K \\
with $B$ tags                              &   &  &   & 1.7K & 3.4K & 34K \\
\hline
$B\rightarrow K^{*}\gamma$ &  &  &  &   &    &  $\star$  \\
(3-8)$\times\,10^{-5}$ & 170/-   &27K/24K  &   & 6K  & 12K & 120K \\
$\delta \, (A_{{\it CP}})$  &   & 0.01 &   & 0.02 & 0.01 & $<$0.01 \\
\hline
$B\rightarrow X_{s}\nu\bar{\nu}$ &  &  &  &   &    &  $\star$  \\
(4.1$\pm$0.9)$\times\,10^{-5}$ &   &  &   & 8  & 16 & 160 \\
$B\rightarrow K^{*}\nu\bar{\nu}$ &  &  &  &   &    &  $\star$  \\
(5$\times\,10^{-6}$) &   &  &   & 1.5 & 3 & 30 \\
\hline
$B\rightarrow X_{s}\mu^{+}\mu^{-}$ & & $\star$  &  &   &    &    \\
(6.0$\pm$1.5)$\times\,10^{-6}$ &    & 7.2K/- &   & 300  & 600 & 6K \\
$B\rightarrow X_{s}e^{+}e^{-}$ &   & $\star$  &  &   &    &    \\
(6.0$\pm$1.5)$\times\,10^{-6}$ &    & 7.2K/- &   & 350  & 700 & 7K \\
$B\rightarrow K^{*}\mu^{+}\mu^{-}$ &  & $\star$  & $\star$  &   &    &  $\star$  \\
(2$\pm$1 $\times\,10^{-6}$) & 61/60-150  & 4.4K/4.5K  & 665/4.2K  & 
120 & 240 & 2.5K \\
$B\rightarrow K^{*}e^{+}e^{-}$ &  &  &  & 150  & 300   &  3K  \\
(2$\pm$1 $\times\,10^{-6}$)    &   &  & &  &   & \\
\hline
$B^{o}_{d}\rightarrow \tau^{+}\tau^{-}$ &  &  &  &   & $\star$   & $\star$  \\
(10$^{-7}$)  &    &   &   & $<$10$^{-5}$  &  $<$2$\times$10$^{-6}$  &
 $<$10$^{-6}$ \\
\hline
$B\rightarrow \mu^{+}\mu^{-}$ &  &$\star$  & $\star$ &   &    &    \\
~~$B_{s}$ (10$^{-9}$) &  5/1.5-6 & 10/11 &  9/7 &  &  &   \\
~~$B_{d}$ (8$\times $10$^{-11}$) &  0/0 & 2/2 &  0.7/20 &
$<$10$^{-8}$  &  $<$5$\times$10$^{-9}$  &  $<$10$^{-9}$ \\
$B^{o}_{d}\rightarrow e^{+}e^{-}$ &  &  &  &   & $\star$   &  $\star$  \\
(10$^{-14}$)  &    &   &   & $<$10$^{-8}$  &  $<$5$\times$10$^{-9}$  &
 $<$10$^{-9}$ \\
\hline
$B \rightarrow \tau\nu$ &  &  &  &  &  &  $\star$ \\
(5$\times$10$^{-5}$)   &  &  &  &  17 &  34 &  350 \\
$B \rightarrow \mu\nu$ &  &  &  &  &  &  $\star$ \\
(1.6$\times$10$^{-7}$)   &  &  &  &  8 &  16 &  150 \\
\hline
$B^{o}\rightarrow \gamma \gamma$ &   &  &  &   &   &  $\star$ \\
(10$^{-8}$)   &   &   &   &  0.4 &  0.8   &  8 \\
\hline
\hline
\end{tabular}
\end{center}
\end{table*}

For both the hadron collider experiments and Super\babar, 
we assume the machine can achieve the
desired luminosity, which is reasonably assured for the hadron
colliders but less certain for the Super \bm-Factory, where design has just 
begun and there are many technology and accelerator issues.

With these caveats, Table~\ref{tab:comp_pi_pi} compares the rate 
of  tagged $B^{o}\rightarrow \pi^{+}\pi^{-}$ obtained in one year from
Super\babar\ and BTeV. Table~\ref{tab:comp_B_DK} shows the number of 
tagged $B^{+}\rightarrow D^{o}K^{+}, D^{o}\rightarrow  K^{+}\pi^{-}$ in the
two experiments. A comparison of BTeV, LHCb, \babar\ and Belle in 2005,
and the $e^{+}e^{-}$ machines at 10$^{35}$ and 10$^{36}$ is given in 
Table~\ref{tab:comp_e_had_cp} for several states of importance to the
study of {\it CP} violation in $B$ decays. Finally, Table~\ref{tab:comp_e_had_rareb}
shows a comparison of CDF/D0, BTeV/LHCb, ATLAS/CMS, \babar/Belle,
and  $e^{+}e^{-}$ machines at 10$^{35}$ and 10$^{36}$ for ``rare decays''
of the $B$ mesons.

It is clear that the 10$^{36}$ $e^{+}e^{-}$ machine can compete with
the hadron collider experiments on many interesting {\it CP} violating decays
and on rare decays of $B_{d}$  and $B_{u}$.
It should do better on decays involving $\tau$'s and missing
$\nu$'s since the hermeticity and energy constraints provided by running
at threshold permit one to establish the neutrino's presence in the event
by demonstrating a recoil mass consistent with zero. While 
$B^{o}\rightarrow \pi^{o} \pi^{o}$ may be barely detectable in several
years of operation at  the 10$^{36}$ $e^{+}e^{-}$ machine, none of the
hadron experiments have yet claimed to be able to observe this state.

The tables are designed to compare the $e^{+}e^{-}$ machines with the
hadron machines in the areas where the former are strong. To have a complete
picture, one needs to  remember that the  $e^{+}e^{-}$ machine can do only
very limited $B_{s}$ physics compared to the hadron collider experiments.
In particular, the proper time resolution, $\sigma_{\tau}$ of 900
$f{\rm s}$, compared to better than 40 $f{\rm s}$ for BTeV, and LHCb, 
precludes the
study of time dependent effects in $B_{s}$ decays. This is a strength of
the hadron collider experiments. The  $e^{+}e^{-}$ experiments also
do not have high enough energy to study $b$-baryons or $B_{c}$ mesons.

\section{Giga-$Z$ machines}

The LEP experiments, running on the $Z$, were able to make many
important \bm\ physics measurements even though the luminosity was
only $\sim 10^{31}{\rm cm}^{-2}{\rm s}^{-1}$. 
SLD, by exploiting the ability to polarize the electron
beam at a linear collider, was able to make significant measurements
at an even lower luminosity. As plans develop to build a high energy,
high luminosity $e^{+}e^{-}$ linear collider, it is worth considering
whether competitive \bm\ physics at the $Z$ can be carried out at these 
facilities~\cite{Gudrun}\cite{Klaus}.

The reasons why the $Z$-pole is a good place to study $B$ physics are:
\begin{itemize}
\item The cross section for producing states containing $b$-quarks is
large, $\sim 6.6$nb;
\item The signal to background is very favorable, $\sim 25$\%;
\item All species of $b$-hadrons are produced, including $B_{s}$
      and $\Lambda_{b}$;
\item The $B$'s have a large boost so that time-evolution studies are possible;
\item Due to the high boost, the two $b$-hadrons are well separated and
separated from the interaction vertex; and
\item The beams can be polarized. This leads to a correlation
between $b$-direction, and the $B$ hadron direction, 
with respect to the $e^{-}$ direction,
which constitutes a highly efficient flavor tag. Electron polarizations
of $>$80\% are achievable and it is expected that positron polarizations
of $\sim$60\% can be obtained.
\end{itemize}
Even though the attainable $b$ yield is low compared to the
hadron colliders or Super\babar, these features permit
the extraction of clean, tagged samples with very high
efficiency, since all $B$'s are triggered and reconstructed and tagging
is very efficient.  The high efficiency  partially offsets the low produced 
rates.

Typical design luminosities for an $e^{+}e^{-}$ linear collider designed
to run at 500 GeV center of mass energy are 2-3$\times 10^{34}$. As part of 
the program of electroweak physics studies that can be done at these
machines, there will be some running at the $Z$, in order to make better
measurements of electroweak parameters and to make rigorous
tests of the consistency of the Standard Model. It seems to be currently
accepted that a run that produces $10^{9}$ $Z$'s is what is required.
At that level of statistics, some measurements are already limited
by the understanding of how to make theory corrections while others
are limited by the experimental systematic errors, for example
in measuring the polarization or the center of mass energy.
  
Even with the lower luminosity, say $5\times 10^{33}$, expected at the $Z$,
it would take only 50 days to accumulate $10^{9}$ $Z$'s with polarization
of 0.8 for electrons and 0.6 for positrons. This provides a sample
of $\sim 4\times 10^{8}$ $b$-hadrons for studies.

There are plans for a dedicated $Z$ facility associated with the
high energy collider. Based on the remarks on electroweak physics, \bm\
physics would have to provide the justification for this. The objective would
be to achieve $10^{10}$ $Z$'s, corresponding to $\sim 4\times 10^{9}$
$B$-hadrons. Table~\ref{tab:giga-z} compares the $\sin 2\beta$ reach for
this facility with the \bm-factories and the hadron collider experiments.
It is clear that even $10^{10}$ $Z$'s, which takes 3-5 years to obtain, is 
barely competitive with one year of data from BTeV/LHCb or Super\babar.

This, however, is not the entire story. There are several classes of
studies that take advantage of the unique characteristics of 
$b$-quark production at the $Z$. These include:
\begin{itemize}
\item States that are polarized, especially $b$-baryons;
\item Searches for direct {\it CP} asymmetries in rare decays, such
as $b\rightarrow s\gamma$ and $b\rightarrow s l^{+}l^{-}$;
\item Measurements involving inclusive final states;
\item ``Missing Energy'' modes, such as $b\rightarrow s \nu \bar{\nu}$
and $B\rightarrow \tau\nu$; and
\item Rare $Z\rightarrow b\bar{s}\,+\,\bar{b}s$ which are expected
to be too small to observe in the Standard Model.
\end{itemize}
These classes of decays might reveal new physics.

Polarization studies are a case in point. The $b$ quarks are strongly 
polarized. It is a prediction of HQET, confirmed by experiment, that the
polarization survives the hadronization process. OPAL has measured
\begin{eqnarray}
P_{\Lambda_{b}} & = & -0.56^{+0.20}_{-0.13}\pm0.09 
\end{eqnarray}
Thus, the Giga-$Z$ facility can be viewed as a 
high luminosity, $\sim 10^{8}$/year source of polarized $\Lambda_{b}$'s.
A study of the angular correlation in $\Lambda_b \rightarrow \Lambda \gamma $ 
\cite{Hiller:2001zj} between the photon direction
and the spin of the $\Lambda_{b}$ is sensitive to spin-flip effects due
to New Physics beyond the Standard Model. In particular, enlarged
spin-flip contributions can be sizeable in L-R symmetric models or SUSY
models with flavor non-universal breaking. The hadronic rare decay
$\Lambda_{b}\rightarrow \Lambda \phi$ also is a probe of New Physics,
although it is theoretically less clean. Table~\ref{tab:b-bary-decays}
gives a list of potentially interesting decays modes. There are many other
interesting topics in $b$-baryon physics that can be explored.

The case for a dedicated  Giga-$Z$ facility at the $Z$ in a future $e^{+}e^{-}$
linear collider is just beginning to be discussed and  needs much more 
development followed by a careful assessment of the contributions it can 
make to the picture of rare $B$ decays and {\it CP} violation.

\begin{table}
\begin{center}
\caption[]
{Comparison of the $\sin 2\beta$ reach with $10^{9}$ and $10^{10}$ $Z$'s}
\label{tab:giga-z}
\begin{tabular}{l|c|c|c|c} \hline
                     & $e^{+}e^{-}$(2005) & BTeV/LHCb &
$10^{9}$ $Z$  & $10^{10}$ $Z$  \\
                     &                    & $10^{7}$s &  & (3-5 yrs) \\
\hline
$\delta \sin 2\beta$ & 0.037& 0.014/0.02  & 0.04 &  0.013  \\ \hline
\end{tabular}
\end{center}
\end{table}

\begin{table}
\begin{center}
\caption[]
{Interesting $b$-baryon decay modes which can be studied at the $Z$.}
\label{tab:b-bary-decays}
\begin{tabular}{ll} \hline
\multicolumn{2}{l}{Semileptonic:} \\
                  & $\Lambda_{b}\rightarrow \Lambda_{c}l\nu_{l}$ \\
                  & $\Lambda_{b}\rightarrow pl\nu_{l}$ \\ \hline 
\multicolumn{2}{l}{rare:} \\                
\multicolumn{2}{l}{~~radiative:} \\                
                  & $\Lambda_{b}\rightarrow \Lambda \gamma$ \\
\multicolumn{2}{l}{~~semileptonic:} \\                
                  & $\Lambda_{b}\rightarrow \Lambda l^{+}l^{-}$ \\
                  & $\Lambda_{b}\rightarrow \Lambda \nu\bar{\nu}$ \\ \hline
\multicolumn{2}{l}{hadronic:} \\                
                  & $\Lambda_{b}\rightarrow \Lambda \phi$ \\
                  & $\Lambda_{b}\rightarrow n D^{*o}_{2}$ \\ \hline
\multicolumn{2}{l}{inclusive:} \\                
                  & $\Lambda_{b}\rightarrow X_{s} \gamma$ \\ \hline
\end{tabular}
\end{center}
\end{table}

\section{Conclusion}

$\lowercase{\phi}$ Factories have a broad program with many unique and 
desirable features, but, in the area of rare kaon decays, they are
unlikely to have sufficient flux to challenge the dedicated Fixed Target 
experiments.

The PEP-N physics program is well-defined, unique and timely. 
This is especially true of
the measurement of R. However, there was no clear demonstration at Snowmass 
that the required systematic error per point (about 2\%) could be achieved. 
Control of systematic 
errors needs to be carefully evaluated before proceeding with PEP-N.

CESR-c/CLEO-c promises a 400-fold increase in $D$ meson data at threshold. 
The data would provide 
a crucial and timely validation of lattice QCD, HQET, ChPTHH and other theoretical 
techniques which are central to progress in flavor physics in this decade, and in the case of  
lattice QCD, also a key to  
addressing strong coupling that may be a feature of the physics beyond the 
Standard Model that 
we expect to be discovered at the LHC. CLEO-c also promises (a) A factor 4-12 improvement 
in key hadronic branching ratios which will set the absolute scale for beauty and charm 
quark physics. (b) A significant improvement, ($\times 5-10$) in CKM matrix element precision 
in the charm sector, and ($\times 2-8$) 
in the beauty sector in conjunction with data obtained at experiments 
with a \bm\ physics capability 
at $e^+e^-$ \bm-factories and hadron colliders. (c) CLEO-c has discovery potential, since 
the experiment is sensitive to new physics through $D$ mixing, 
$D$ {\it CP} violation and rare decays  of $D$ mesons and 
the $\tau$ lepton, and in the search for new forms of matter, including 
glueballs and hybrids.
Finally a flexible accelerator, an experienced collaboration and a high quality detector 
are already in place, making the well-defined three year physics program very attractive.

BES/BEPC is currently proposing to upgrade the machine and detector.
BEPC II would be a two
ring machine with a peak luminosity in excess of 
$ 10^{33} {\rm cm}^{-2}{\rm s}^{-1} $. A completely new detector, BES III,
would be built. BEPCII/BESIII would come on line around 2006
and would accumulate a data sample
one order of magnitude larger than CLEO-c. 
The physics program  of BES III is identical to CLEO-c. 
For BES III to make a significant impact it 
is absolutely essential that the detector be as good as the CLEO-c detector.
If that can be achieved, the significantly larger luminosity of BEPCII over CESR-c is likely 
to be a considerable advantage for new physics reach.
For CKM physics, theory will have to sharpen for the larger statistics of BES III
to be used to full advantage. 
Hall D at TJNAL, coming on-line in 2006, and CLEO-c have some overlap
but there is also complementarity. CLEO-c is focusing on glue rich states and vector
hybrids both light and heavy. Hall-D is focused on states with exotic quantum numbers.
There is also a proposal from the GSI accelerator in Germany for a High Energy Storage Ring
(HESR) for antiprotons. The charmonium studies this machine will allow are likely 
to be complementary to CLEO-c.

The two asymmetric \bm-factories, PEP-II and KEKB, have achieved reliable
operation at high luminosities of a few $10^{33}{\rm cm}^{-2}{\rm s}^{-1}$ in
a remarkably short time. Both 
machines have plans for incremental
upgrades which eventually are expected to produce luminosities of
$10^{35}{\rm cm}^{-2}{\rm s}^{-1}$, which corresponds to
$10^{9}$ $B$  pairs per year. These asymmetric super \bm-factories 
have significant
advantages with respect to hadron colliders in terms of
$\pi^{o}$ detection efficiency, $\nu$ reconstruction and generally smaller backgrounds.
In this report, as an example of what can be achieved by a long
run at $10^{35}{\rm cm}^{-2}{\rm s}^{-1}$, we discussed only the super KEKB/Belle 
upgrade. The PEP-II analog has identical physics reach. 
(For PEP, we concentrated  
Super\babar\ with a design luminosity of $10^{36}{\rm cm}^{-2}{\rm s}^{-1}$.) 
The high statistics of a $10^{35}{\rm cm}^{-2}{\rm s}^{-1}$ 
super \bm-factory allows  significant numbers of 
$B$ mesons to be tagged by full reconstruction,
and this permits many significant physics studies to be performed
especially involving final states with a neutrino such as 
semileptonic $b \rightarrow u $ transitions to determine $V_{ub}$,
leptonic decays and electroweak penguins.
The KEKB machine upgrade is believed to be feasible.
Operation at $10^{35}$ will produce significantly higher background rates
in Belle which will lead to
larger occupancy. Accordingly, the detector will need several upgrades 
which we judge to be feasible. After several years of running
at  $10^{35}$, the \bm\ physics program at KEK will probably end. 
A clear consensus was reached in the E2 group that an $e^+e^-$ 
\bm-factory operating at 
$10^{35}$ would not be competitive with experiments at hadron colliders specifically
LHCb/BTeV/ATLAS/CMS coming on-line around 2007. This view is also held 
by the proponents of the KEKB/Belle upgrade.

The Super \bm-factory is a new continuous 
injection $e^+e^-$ collider that would 
operate in the PEP-II tunnel or the SLC arcs
at a luminosity of $10^{36}{\rm cm}^{-2}{\rm s}^{-1}$, 
a factor 300 more than PEP-II achieves today. 
It has been proposed specifically 
to be complementary to the  hadron collider \bm\ experiments as a precision
probe of the consistency of the flavor changing sector of the Standard
Model and in searches for New Physics.  Occupancy and machine backgrounds
will probably require the replacement of the entire \babar\ detector. 
The detector design is challenging, raising many difficult  R\&D issues. 
Assuming detector efficiency
could be maintained at such a high luminosity, 
we estimate that Super\babar\ would be 
complementary to LHCb/BTeV for rare decays of $B_{d}$ and $B_{u}$ mesons, 
superior for decays with $\nu$'s, and competitive for
decays with  a $\pi^{o}$,or $\gamma$. It accuracy would be
comparable for the angles $\alpha$, $\beta$ and $\gamma$ but not $\chi$.    
Compared to hadron collider experiments, the $B_s$ program would be limited 
by the complications of operating at the $\Upsilon(5S)$, and because of
much poorer proper time resolution.
There would be no $\Lambda_b$ or $B_c$ physics. 

The sensitivities of future hadron collider experiments have been determined from
detailed and sophisticated simulations of signals and backgrounds. As these simulations
are an approximation to reality the expected performance of LHCb and BTeV may be
somewhat better or somewhat worse than the simulations predict.
Projections for Super\babar\ are at this point 
mainly done by scaling from \babar\ experience assuming that the new
detector, which still has many open R\&D issues, will achieve the same 
efficiency that \babar\ now achieves even though the luminosity will be a 
factor of 300 higher. More
realistic studies need to be performed before a full comparison 
between Super\babar\ and the hadron collider experiments is made.  
It is also important to quickly implement
a realistic machine lattice and IR design to provide predictions of 
the very large machine backgrounds
that will exist at Super\babar, especially background due to continuous injection.
If backgrounds prove tractable, and detector simulations support the simple scaling from \babar\ experience, an R\&D program on the machine and detector 
should be initiated. 

The case for a dedicated  Giga-$Z$ facility at a future $e^{+}e^{-}$
linear collider is just beginning to be discussed and needs much more 
development followed by a careful assessment of the contributions it can 
make to our understanding of rare \bm\ decays and {\it CP} violation.

In conclusion, $e^+e^-$ colliders at low energy have played an important role
in the development of our understanding of flavor physics, non-peturbative QCD 
and radiative corrections. Today the Fixed Target hadron experiments
appear to be the best way to address
key  measurements in kaon physics involving rare decays. 
Electron positron colliders 
have a unique role in the measuremnt of R, and are complementary to hadron 
colliders as a probes of non-peturbative QCD, and charm and beauty flavor 
physics. The physics is more important than the method used. It would be 
prudent to  carefully evaluate the merits of both hadron colliders 
and $e^+e^-$ colliders for each application at each stage
in our quest, only ruling out one approach when it clearly fails. 
In these areas, competition, complementarity, and even some redundancy
have proven to important to ultimate  progress.

\begin{acknowledgments}

We wish to thank more than fifty of our colleagues who made
very valuable contributions to the E2 working group at Snowmass.

\end{acknowledgments}

\end{document}